# Methane release on Early Mars by atmospheric collapse and atmospheric reinflation

Edwin S. Kite[1], Michael A. Mischna[2], Peter Gao[3,4], Yuk L. Yung[2,5], Martin Turbet[6,7,*]

1. University of Chicago (kite@uchicago.edu), Chicago, IL 60637.
2. Jet Propulsion Laboratory, California Institute of Technology, Pasadena, CA 91109.
3. NASA Ames Research Center, Mountain View, CA 94035.
4. University of California, Berkeley, CA 94720.
5. California Institute of Technology, Pasadena, CA 91125.
6. Laboratoire de Météorologie Dynamique, IPSL, Sorbonne Universités, UPMC Univ Paris 06, CNRS, Paris 75005, France.
7. Observatoire Astronomique de l'Université de Genève, 51 Chemin des Maillettes, 1290 Sauverny, Switzerland.
* Currently Marie Sklodowska-Curie Fellow, Geneva Astronomical Observatory.

## Abstract.

A candidate explanation for Early Mars rivers is atmospheric warming due to surface release of $H_2$ or $CH_4$ gas. However, it remains unknown how much gas could be released in a single event. We model the $CH_4$ release by one mechanism for rapid release of $CH_4$ from clathrate. By modeling how $CH_4$-clathrate release is affected by changes in Mars' obliquity and atmospheric composition, we find that a large fraction of total outgassing from $CH_4$ clathrate occurs following Mars' first prolonged atmospheric collapse. This atmosphere-collapse-initiated $CH_4$-release mechanism has three stages. (1) Rapid collapse of Early Mars' carbon dioxide atmosphere initiates a slower shift of water ice from high ground to the poles. (2) Upon subsequent $CO_2$-atmosphere re-inflation and $CO_2$-greenhouse warming, low-latitude clathrate decomposes and releases methane gas. (3) Methane can then perturb atmospheric chemistry and surface temperature, until photochemical processes destroy the methane.

Within our model, we find that under some circumstances a Titan-like haze layer would be expected to form, consistent with transient deposition of abundant complex abiotic organic matter on the Early Mars surface. We also find that this $CH_4$-release mechanism can warm Early Mars, but special circumstances are required in order to uncork $10^{17}$ kg of $CH_4$, the minimum needed for strong warming. Specifically, strong warming only occurs when the fraction of the hydrate stability zone that is initially occupied by clathrate exceeds 10%, and when Mars' first prolonged atmospheric collapse occurs for atmospheric pressure > 1 bar.

## 1. Introduction.

### 1.1. Causes and effects of methane release on Early Mars.

The Mars Science Laboratory's (MSL's) investigation of Early Mars sediments has turned up two results which remain a puzzle. The first puzzle is the discovery of a paleolake (Grotzinger et al. 2014). Paleolakes on Mars individually lasted as long as >($10^2$–$10^3$) yr (e.g., Irwin et al. 2015). Such longevity is hard to reconcile with climate models that predict no lakes, or only short-lived lakes (Wordsworth 2016, Hynek 2016, Luo et al. 2017, Haberle et al. 2017, Vasavada 2017, Kite et al. 2019, Kite 2019). The second puzzle is the detection by MSL of organic matter (Eigenbrode et al. 2018). The organic matter is found in excess of known instrumental backgrounds. The source could be exogenic (asteroidal dust), or indigenous to Mars (for example, magmatic or biological), but remains unknown.

Both these puzzles – lake-forming climates and organic matter production - could, in principle, be explained by $CH_4$ release. Methane is a greenhouse gas. For one-dimensional calculations of Early Mars climate, $CH_4$–induced warming (due to $CH_4$-$CO_2$ collision-induced absorption) overwhelms $CH_4$–induced cooling (due to $CH_4$ absorption of sunlight) when $pCO_2$ exceeds 0.5-1.0 bar (Wordsworth et al. 2017, Turbet et al. 2019). Indeed, $CH_4$ bursts have been proposed as a cause of early Mars lake-forming climates (Chassefière et al. 2016, Kite et al. 2017a), and as a cause of various Early Mars geologic features (e.g., Skinner & Tanaka 2007, Kite et al. 2007, Oehler & Allen 2010, Wray & Ehlmann 2011, Komatsu et al. 2016, Ivanov et al. 2014, Oehler & Etiope 2017, Pan & Ehlmann 2014, Etiope & Oehler 2019, Brož et al. 2019). Detections of tiny amounts of $CH_4$ in Mars' present-day atmosphere have been reported (e.g. Webster et al. 2018), although these detections do not all agree with one another and are in tension with models that predict lower methane abundance, and also lower methane abundance variability, than has been reported (Zahnle et al. 2011, Mischna et al. 2011, Korablev et al. 2019, Zahnle & Catling 2019, Moores et al. 2019). Separate from its potential to warm Mars, methane can also serve as the feedstock for photochemical production of complex organic matter – hazes and soots. Haze-derived soot on Titan today, which is ultimately derived from $CH_4$, accumulates on Titan's surface (Lunine & Atreya 2008, Hörst 2017). A ratio of $CH_4/CO_2$ > 0.1 is needed for vigorous haze production (Trainer et al. 2006, Haqq-Misra et al. 2008, Kite et al. 2017a).

Here we investigate a new scenario for Mars methane release (Fig. 2). The scenario involves both collapse of the $CO_2$ atmosphere, and also the reverse process – $CO_2$-atmosphere reinflation (Gierasch & Toon 1973, Soto et al. 2015, Wordsworth et al. 2015). Atmospheric collapse on Mars shifts the climate from a state with all $CO_2$ in the atmosphere, to a state with most $CO_2$ in the form of surface ice, and atmospheric reinflation is the reverse process. In our model, atmospheric-collapse-triggered depressurization of $CH_4$ clathrate and subsequent reinflation-induced warming of $CH_4$ clathrate causes transient outgassing of $CH_4$.

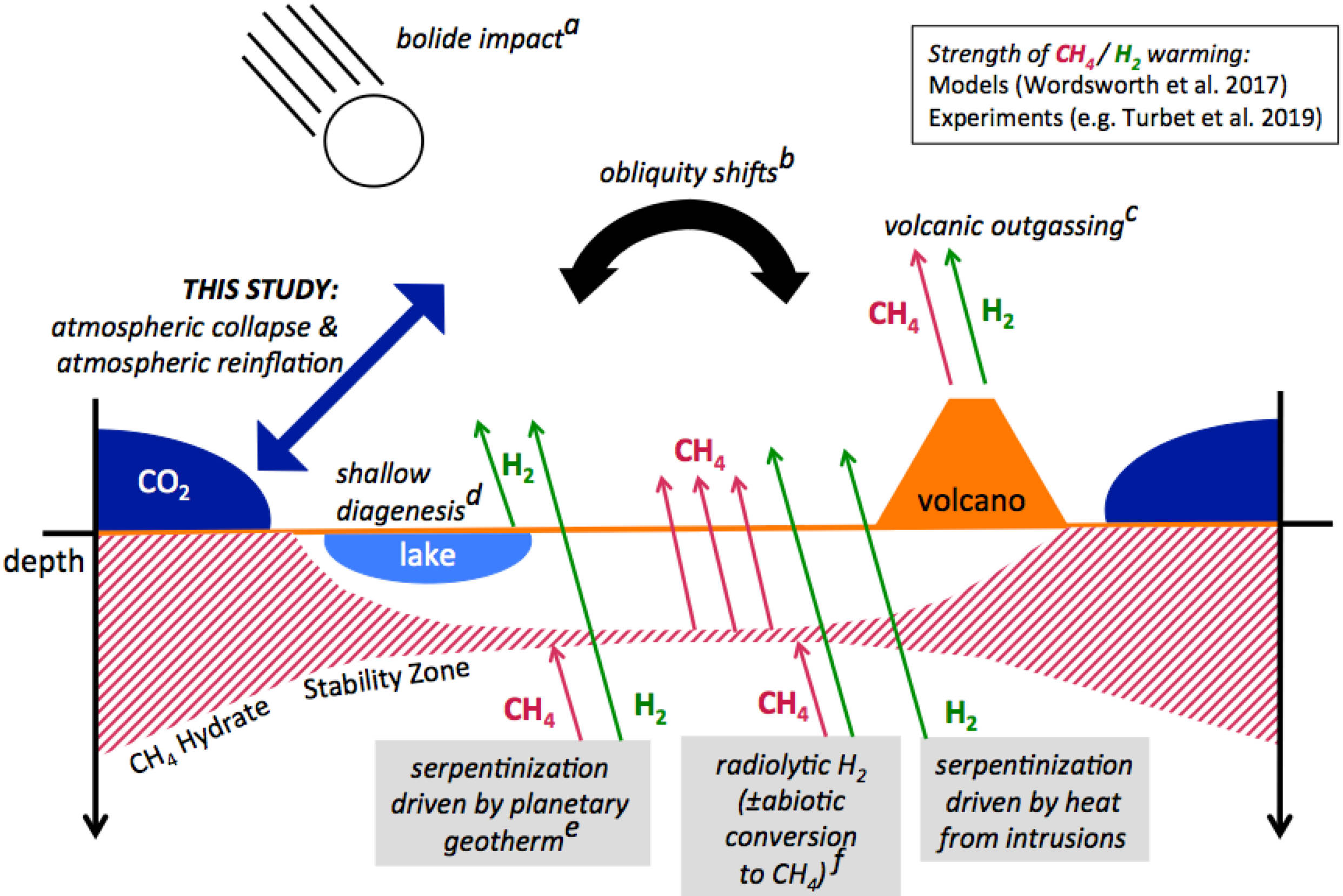

Figure notes: *a*: e.g. Haberle et al. 2018, Steakley et al. 2019. *b*: e.g., Prieto-Ballesteros et al. 2006, Kite et al. 2017a. *c*: e.g., Batalha et al. 2016, Ramirez 2017. *d*: Tosca et al. 2018. *e:* e.g. Chassèfiere et al. 2016. *f:* e.g. Lin et al. 2005, Tarnas et al. 2018.

**Fig. 1.** This paper in context: candidate drivers and sources for transient release of $H_2$ or $CH_4$ gas on Early Mars. A major effort, involving many research groups, is underway to determine the conditions (if any) under which transient reducing greenhouse atmospheres on Mars are geologically plausible, by testing each proposed mechanism. One chemical reaction that could link candidate driver *a* to transient release of $H_2$ is reaction between Fe from the impactor and H2O from Mars (i.e., $Fe + H_2O \rightarrow FeO + H_2$).

In order to model these processes, our approach combines the following elements (see Appendix A for details): climate models, obliquity calculations, a parameterization of the photochemical destruction of $CH_4$ that is fit to a detailed photochemical model, and a model of the (latitude/longitude/ depth-dependent) stability of $CH_4$ clathrate. For simplicity we omit other clathrates such as $CO_2$ clathrate (and $CH_4$-$CO_2$ clathrate hybridization; Sloan & Koh 2008). We emphasise $CH_4$ clathrate for the following reasons. Methane clathrates are a candidate driver of past climate change on Titan and Earth (e.g., Tobie et al. 2006, Bowen et al. 2016). Moreover, $CH_4$ storage in clathrates has been previously proposed for Mars (Chassefière et al. 2016, and references therein), and as a possible source of the $CH_4$ reported by a team using MSL Sample Analysis at Mars (SAM) Tunable Laser Spectrometer (TLS) data (Webster et al. 2018; but see also Zahnle et al. 2011 and Zahnle & Catling 2019). However, $CH_4$ has a $<10^6$ yr photochemical lifetime (Sagan 1977, Gierasch & Toon 1973, Chassefière et al. 2016, Lasue et al. 2015, Wordsworth et al. 2017, Kite et al. 2017a). In order to build up the high (≥1%) levels of atmospheric $CH_4$ that are needed for strong warming on Early Mars, $CH_4$ supply must overwhelm photolysis, and so outgassing must be swift. In turn, swift outgassing requires a mechanism that (a) efficiently traps $CH_4$, yet (b) subsequently releases the $CH_4$, (c) in large quantities (d) at a rate that outpaces photochemical destruction. Here, we show that a suitably vigorous release mechanism is atmospheric collapse and atmospheric reinflation.

## 1.2. Previous work on Mars atmospheric collapse and atmospheric reinflation.

$CO_2$-atmosphere collapse (a runaway shift from intermediate to very low $P_{CO2}$) is distinct from $CO_2$-atmosphere condensation (which curtails $P_{CO2}$ at high pressure, ≳3 bars, without involving a runaway shift). The most recent 3D analysis of runaway atmospheric collapse on Mars is by Soto et al. (2015). In their study, which assumed present-day solar luminosity, collapse onto Olympus Mons plays a key role. However, this volcano was probably not very tall at 3.7 Ga (Isherwood et al. 2013). Idealized models have also been used to study atmospheric collapse on Mars (Nakamura & Tajika 2001, 2002, 2003), and on exoplanets (e.g., Heng & Kopparla 2012, Wordsworth 2015, Turbet et al. 2017). All models agree that, for a given $P_{CO2}$, the main control on atmospheric collapse is obliquity. The obliquity for atmospheric collapse depends on assumed $CO_2$ ice cloud grain size (Kitzmann 2016). Using a 3D model, Kahre et al. (2013) assessed the importance of carbon dioxide ice cap albedo and emissivity in setting the boundaries of the atmospheric collapse zone. Manning et al. (2006, 2019) considered atmospheric collapse as part of a study of the long-term evolution of the Mars carbon inventory. There is strong geologic evidence for major changes in $P_{CO2}$ due to redistribution of $CO_2$ between the atmosphere and $CO_2$ ice caps even in the geologically recent past, within the last 5 Myr (e.g., Kreslavsky & Head 2005, Phillips et al. 2011, Bierson et al. 2016, Manning et al. 2019)[1]. No previous study links Mars atmospheric collapse and reinflation to a warmer-than-pre-collapse climate (Halevy & Head 2017).

---

[1] New radar constraints on the present-day inventory of $CO_2$ ice (Bierson et al. 2016, Putzig et al. 2018) mean that warming driven by the direct greenhouse effect of $CO_2$ released into the atmosphere during re-inflation in the geologically recent past or near future is probably small (Jakosky & Edwards 2018). This is contrary to the large warming envisaged by early studies (Sagan et al. 1973, Gierasch & Toon 1973,

## 2. Setting the stage for atmospheric-collapse-initiated $CH_4$ release.

The atmospheric-collapse-initiated methane release scenario is motivated by evidence for the following:- (1) Noachian-age water-rock reactions in the deep subsurface; (2) a ≳0.5 bar past atmosphere that (3) drove surface $H_2O$ ice to high ground; and (4) large-amplitude obliquity variations. Understanding of all four has improved in recent years, as we now discuss.

1) Noachian-age hydrothermal minerals record Noachian hydrothermal reactions (e.g., Carter et al. 2013, Ehlmann et al. 2010, Parmentier & Zuber 2007, Sun & Milliken 2015). Hydrothermal reactions between Mars' mafic/ultramafic crust and waters charged with magmatic and/or atmospheric C should yield both $H_2$ and $CH_4$ (e.g., Lyons et al. 2005, Oze & Sharma 2005, Klein et al. 2019). $CH_4$ is emphasized here (Etiope & Sherwood Lollar 2013), due to its stability in clathrate hydrates under lower pressures and higher temperatures relative to $H_2$. Although abiotic reactions can, in principle (stoichiometry + thermodynamics), form ~$10^2$ bars of $CH_4$, $CH_4$ production by abiotic reactions faces kinetic barriers (e.g., Oze et al. 2012, Tarnas et al. 2019). These barriers can be overcome by high temperatures in olivine-rich rocks that host catalysts (e.g., Ni; Etiope & Schoell 2014), and by recycling of fluid by hydrothermal circulation. Both hydrothermal circulation, the hydrothermal reaction-rate, and radiolytic $H_2$ production slackened as the Mars geotherm cooled and Mars' radionuclide abundances decreased through radioactive decay (Ehlmann et al. 2011, Fassett & Head 2011, Tarnas et al. 2018). Hydrothermal circulation would transport $CH_4$ that was produced in the deep subsurface up to the cooling near-surface (e.g., beneath ice sheets or primordial seas), where it would have been trapped as $CH_4$ clathrate hydrate (Fig. 2) (Chastain & Chevrier 2007, Chassefière & Leblanc 2011, Mousis et al. 2013). $CH_4$-clathrate is not stable at Mars' surface (Fig. 4), but is stable in the subsurface. Specifically, $CH_4$-clathrate is stabilized by increasing pressure (stable for $P \geq 21$ bar at 273.15 K, corresponding to a depth of 400m within 1.5 g/cc regolith), and by decreasing temperature (stable for $P \geq 0.015$ bar at 200 K) (Fig. 4). Shallow-subsurface $CH_4$ clathrate, once formed, could later be destabilized by orbital forcing (Kite et al. 2017a, Prieto-Ballesteros et al. 2006) or by other mechanisms (Chassefière et al. 2016; this paper) (Fig. 1).

---

McKay et al. 1991). Nevertheless, earlier in Mars history, Mars had more $CO_2$, and so the warming driven by the direct greenhouse effect of $CO_2$ released into the atmosphere during re-inflation early in Mars history could be much larger (e.g., the effect of going from 0.01 → 1 bar $P_{CO2}$), as is assumed in this study.

2) Today, $P_{CO2}$ = 6 mbar, too low to permit extensive liquid water (Hecht 2002). Past $P_{CO2}$ was greater, due to gradual atmospheric loss by hard-to-reverse processes (carbonate formation, polar basal melting and infiltration of $CO_2$) as well as by irreversible escape to space (Kurahashi-Nakamura & Tajika 2006, Lammer et al. 2013, Jakosky et al. 2018, Manning et al. 2019). Most estimates of $P_{CO2}$ at the time of the valley networks are in the range 0.2-2 bar (Kite et al. 2014; Kurokawa et al. 2017; Cassata et al. 2012; Warren et al. 2019; Kite 2019 and references therein). Such thick $CO_2$ atmospheres ensure that cold ground is at high elevations (Wordsworth et al. 2013, Wordsworth et al. 2015).
3) Cold ground at high elevations, including most of the Southern hemisphere, would act as a cold trap for $H_2O$ ice (Wordsworth et al. 2013). Average $H_2O$ ice sheet thickness, assuming ice above +1 km elevation, was ≥300 m (e.g., Fastook & Head 2015, Mahaffy et al. 2015). Three hundred meters of $H_2O$ ice on Mars is sufficient to stabilize $CH_4$ clathrate in the regolith pore space beneath the ice sheet.
4) Unlike Earth, Mars is thought to undergo chaotic large-amplitude shifts in spin-orbit parameters; among these variations, the most important are variations in obliquity, $\phi$. $\phi$ varies in the range $0° < \phi < 70°$; Touma & Wisdom 1993, Laskar et al. 2004, Holo et al. 2018), on which are superposed quasi-periodic $\phi$ variations with period $10^5$-$10^6$ yr and amplitude 0-20° (10× Earth).

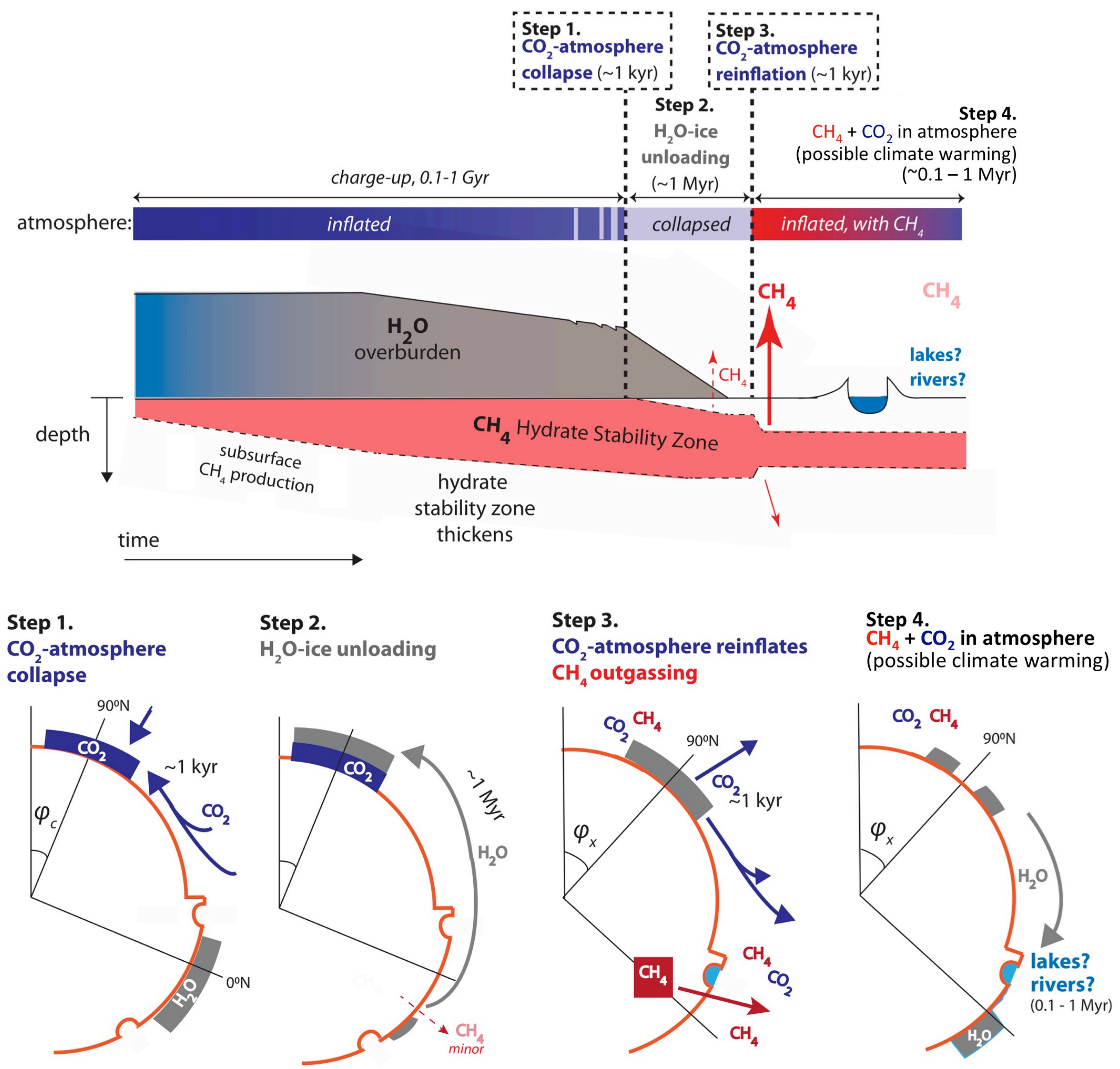

**Fig. 2.** Overview of the collapse-initiated methane release mechanism. *Upper panel:* Depth-versus-time schematic of the long-term evolution of a column of the low-latitude highlands of Mars. Figs. 3-5 show details. *Lower panel:* Latitude-versus-time schematic, showing key volatile reservoirs: $CO_2$, $H_2O$, and $CH_4$. Reservoir thicknesses are not to scale. Below the critical obliquity for atmospheric collapse ($\phi_c$), 90-99.9% of the atmosphere will condense in ~1 kyr (Step 1). This unloads high ground (Step 2), releasing some $CH_4$ from sub-ice clathrate. Later re-inflation of the atmosphere triggered by $\phi$ rise (Step 3) leads to massive $CH_4$ release and possible climate warming (Step 4). The basis for the timescale estimates is GCM output, as described in the text.

## 3. How atmospheric collapse can initiate methane release.

Here, we summarize the collapse-initiated $CH_4$-release scenario (a more detailed description is given in §3.1). In the atmospheric-collapse-initiated $CH_4$-release scenario (Fig. 2), atmospheric collapse (Step 1) causes low-latitude surface water ice to sublimate away. This sublimation loss occurs because atmospheric collapse occurs at low obliquity. At low obliquity and low (collapsed) $pCO_2$, the stable location for water ice is at the poles (as on Mars today). Loss of water ice overburden depressurizes and thus destabilizes $CH_4$ clathrate in low-latitude sub-glacial pore space (Step 2). Subsequent atmospheric re-inflation at high obliquity leads to $CO_2$-induced warming that further destabilizes $CH_4$ clathrate (Steps 3/4). If (and only if) the atmospheric collapse is initiated at $pCO_2$ > (0.5-1.0) bar, then $CH_4$ release can cause surface warming. The potential for surface warming is greatest at Step 4. The $CH_4$ pulse is brought to a close by photochemical destruction of $CH_4$. We proceed through each of these steps in greater detail in §3.1.

The collapse-initiated $CH_4$ release mechanism (Fig. 2) is motivated by two separations of timescales:-

1. $CO_2$-atmosphere collapse and re-inflation takes ~$10^3$ yr (Soto et al. 2015). This ~1 mbar/yr pace is set by the need to radiate away (or supply) the latent heat of sublimation for $CO_2$ ice. Both $CO_2$ collapse and $CO_2$ migration are slow relative to $H_2O$ ice migration ($10^5$-$10^7$ yr, according to our GCM; Appendix). ($H_2O$ ice migration is slow because surface water ice is cold on Early Mars, cold ice has a low saturation vapor pressure, and as a result there is not much water vapor in the atmosphere for winds to transport). Since $H_2O$ ice overburden pressure is important to $CH_4$ clathrate stability, this separation of timescales allows for $H_2O$ ice to be "out of position" when the atmosphere re-inflates. As a result, after reinflation, $CH_4$ clathrate is at the same temperature, but a lower pressure, than in the pre-atmospheric-collapse state. Because $CH_4$ clathrate is destabilized at lower pressure (and destabilized at increasing temperature), $CH_4$ clathrate is destabilized after reinflation relative to pre-collapse conditions.

2. The time lag between surface warming and release of $CH_4$ to the atmosphere is the sum of time for the subsurface to be warmed by thermal conduction, the decomposition time for unstable clathrate (which is fast; Gainey & Elwood Madden 2012), and the time lag between clathrate breakdown and release of $CH_4$ to the atmosphere. This sums to $\lesssim 10^3$ yr, much less than the timescale for photochemical destruction of $CH_4$ (~$10^5$ yr). Thus, $CH_4$ can build up to high levels before decaying.

### 3.1. Steps in the scenario.

Step 1. Collapse of an initially thick $CO_2$ atmosphere. We assume an initially thick ($P_{CO2}$ > 1 bar) atmosphere. As the atmosphere is slowly thinned by escape to space and other processes, the remaining atmospheric $CO_2$ becomes increasingly vulnerable to collapse (Fig. 3). Collapse is triggered if polar $CO_2$ ice caps undergo year-on-year growth. Growth of polar $CO_2$ ice caps occurs below a critical polar temperature. Polar temperature is set by insolation, which increases with obliquity, and by heating from the $CO_2$ atmosphere (both the greenhouse effect and equator-to-pole heat transport increase with $P_{CO2}$). Given secular atmospheric-pressure decline, obliquity variations will eventually lower insolation below the threshold for perennial $CO_2$-ice caps (e.g., Forget et al. 2013). Once year-on-year cap growth begins, the caps trap 90-99% of the atmospheric $CO_2$ within $10^3$-$10^4$ yrs (Soto et al. 2015). Why is this collapse so rapid, and (almost) complete? The cause of the rapid runaway is the positive feedback between polar cooling and ice-cap sequestration of $CO_2$. Post-collapse, $P_{CO2}$ is in equilibrium with polar $CO_2$ ice cap surface temperature - i.e., it is very low (Fig. 3) (Sagan et al. 1973). $CO_2$ ice cap thickness, assuming caps poleward of 80°, is ~1200 m/bar $CO_2$ (or lower if the caps flow; Mellon 1996).

Low-obliquity atmospheric collapse involves a hysteresis (Fig. 3, Fig. A2), but is ultimately reversible. Eventually, obliquity rise will trigger rapid re-sublimation of $CO_2$ condensed at low obliquity. Thus, $CO_2$ trapped as $CO_2$ ice is only sequestered temporarily.

Theory predicts that collapse/re-inflation cycles should have occurred in Mars history. There is sedimentological evidence for $P_{CO2}$ < 10 mbar at ~3.7 Ga (Lapôtre et al. 2016), in which case the first collapse occurred >3.7 Ga. There is also indirect evidence from isotopes in Mars meteorite ALH 84001 that collapse (i.e, a very thin atmosphere) did not occur prior to ~4 Ga (Kurokawa et al. 2017).

$P_{CO2}$ reduction leads to loss of most of the greenhouse effect. The corresponding surface cooling (tens of K) helps to stabilize $CH_4$-clathrate, far outweighing the destabilizing effect of loss of the weight of the atmosphere. Therefore, no $CH_4$ release occurs at Step 1. Furthermore, cold post-collapse conditions are very unfavorable for liquid $H_2O$ on Early Mars (Hecht 2002, Turbet et al. 2017).

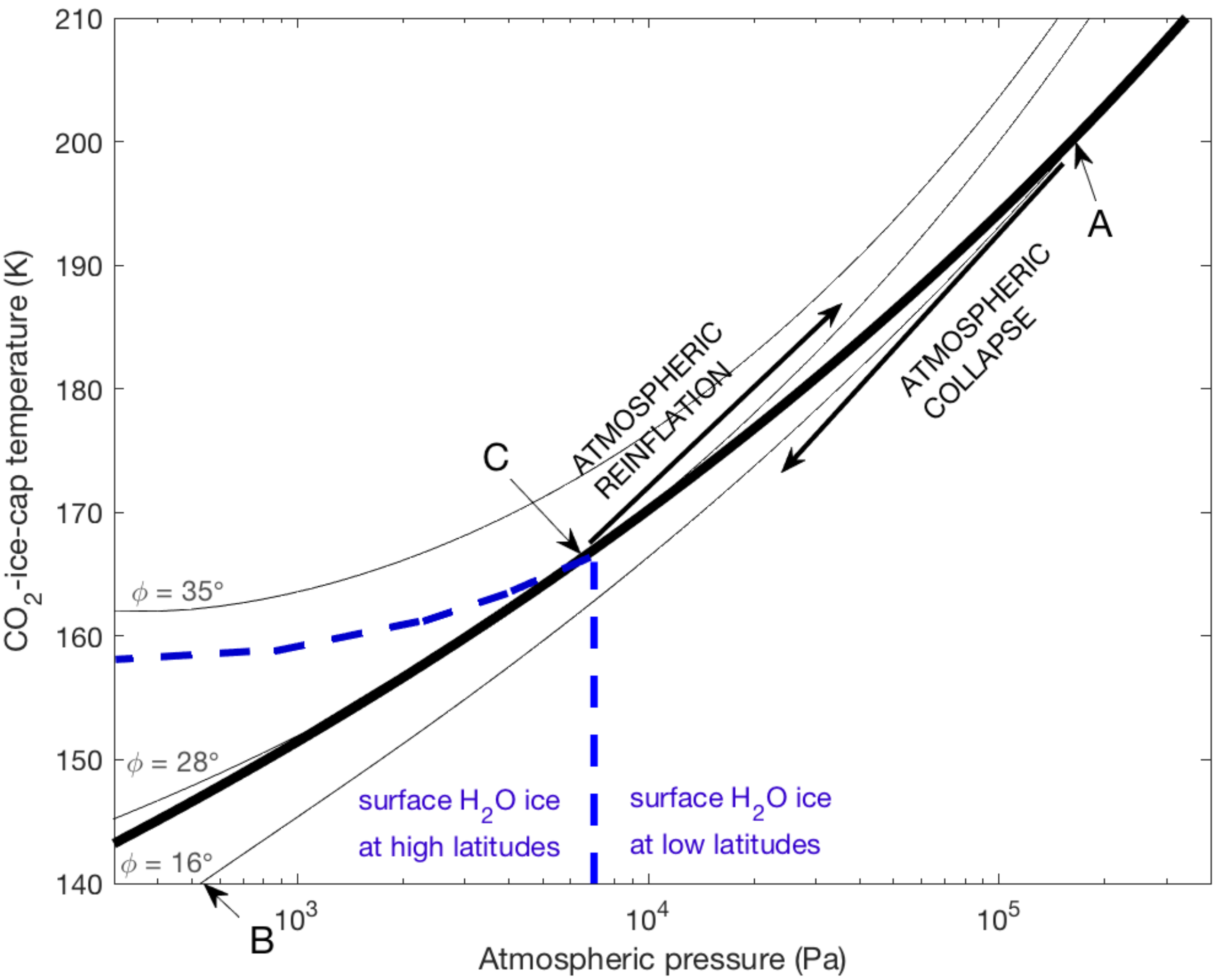

**Fig. 3.** Sketch of atmospheric collapse on Mars, showing how atmospheric collapse drives $H_2O$-ice re-distribution. Thin black lines show annual-mean polar temperature as a function of atmospheric pressure assuming Faint Young Sun insolation (75% of modern insolation). These lines can shift depending on model assumptions, and the lines shown here are illustrative only. Thick black line is the condensation curve for $CO_2$; atmospheres below this line are collapsing onto polar $CO_2$ ice caps (e.g., **A→B**). Blue dashes outline the approximate pressures and obliquities below which surface $H_2O$ ice is stable only at Mars' poles (e.g., Mischna et al. 2013, Wordsworth 2016). Collapse drives a shift of surface $H_2O$ ice from highlands to poles. In this sketch, for an initial $CO_2$ inventory of $8\times10^{18}$ kg (= 2 bar), the atmosphere is stable until obliquity ($\phi$) ≤ 15° (at **A**). Rapid collapse (~$10^3$ yr) moves the system to a very low atmospheric pressure (lower than point **B**). Increasing obliquity (over $10^5$-$10^7$ yr) moves the (ice cap)/atmosphere system along the condensation curve to **C**, (the highest $\phi$ consistent with permanent $CO_2$ ice caps). Further $\phi$ rise leads to sublimation of the $CO_2$ ice cap (~$10^3$ yr), and the system returns to **A.**

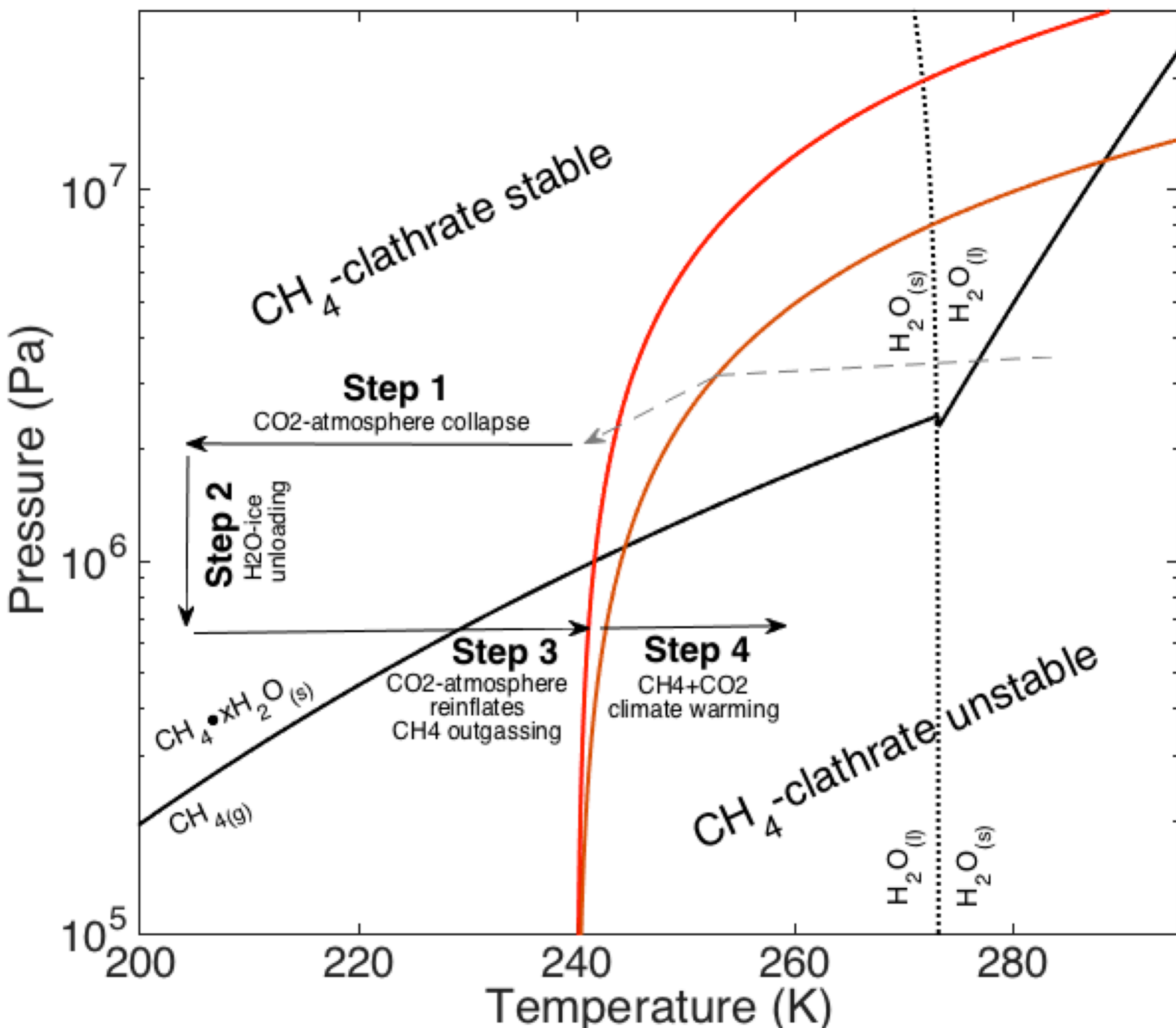

**Fig. 4.** $CH_4$ clathrate phase diagram. Phase boundaries shown in black. The temperatures are obtained from our GCM (which is modified after Mischna et al. 2013; see Appendix for details). Mars geotherms are shown in by upper curved line (early, steep geotherm; red) and lower curved line (later, shallow geotherm; orange). Early in Mars history, near-surface cooling locks-in $CH_4$ as clathrate in regolith beneath ice sheets (dashed gray line with arrow). Further geotherm cooling and escape of ice-sheet $H_2O$ to space has little effect on $CH_4$-clathrate stability. Atmospheric collapse and consequent unloading cause minor $CH_4$-clathrate breakdown (Fig. 2) (Steps 1-2). Warming of the surface upon re-inflation, plus feedback warming, will cause major $CH_4$-clathrate breakdown (Steps 3-4).

Step 2: $H_2O$ ice unloading of low latitudes triggers some $CH_4$ release. Following atmospheric collapse, the collapsed <100 mb atmosphere can no longer supply much heat to the poles. Lacking atmospheric heating, and with obliquity low in the aftermath of atmospheric collapse, the cold poles are now the stable cold-trap location for water ice. Water ice condenses at the poles, and sublimates away from low-latitude highlands. The sublimation rate is slow, due to the Faint Young Sun:

~0.1 mm/yr for water ice albedo = 0.45 (according to our general circulation model, which is a modified version of that in Mischna et al. 2013; Appendix A), rising to ~1 mm/yr for a dust-like ice albedo. Thus, the sublimation timescale for removal of the low latitude $H_2O$-ice depends on the initial $H_2O$ thickness. For a thickness of 300 m, removal/unloading time is $3\times10^5$-$3\times10^6$ yr. On this timescale the absolute distance traveled by $H_2O$ ice via solid-state glacial flow is small (Kite & Hindmarsh 2007, Fastook & Head 2015).

This slow latitudinal shift in $H_2O$ ice overburden pressure depressurizes and thus destabilizes $CH_4$ clathrate in subglacial pore space (Figs. 4-5). Unloading-induced destabilization eventually exceeds cooling-induced stabilization. In other words, the top of the clathrate hydrate stability zone (HSZ) will deepen, so the hydrate stability zone will shrink. The deepening HSZ boundary will cause clathrate stranded above it to decompose in <1 kyr (Stern et al. 2003, Gainey & Elwood Madden 2012). Decomposition will release $CH_4$ (Figs. 4-5). This $CH_4$ is assumed in our model to be swiftly outgassed to the atmosphere, in part because of fracturing induced by the large volume change involved in clathrate decomposition.

The $CH_4$ release to the atmosphere is proportional to the fraction of surface area initially shrouded by $H_2O$-ice. This fraction is ≈ 50% according to the model of Wordsworth et al. (2015)[2]. The initial $CH_{4(g)}$ flux is also proportional to $f$, the fraction of the not-yet-degassed HSZ volume occupied by clathrate. The clathrate is charged up with $CH_4$ produced by water-rock reactions over $>10^8$ yr (Fig. 2a). It is released $<10^3$ yr after destabilization (Stern et al. 2003, Gainey & Elwood Madden 2012), possibly via explosive blow-outs or mud volcanism (Andreassen et al. 2017, Komatsu et al. 2016, Brož et al. 2019).

$P_{CH4}$ on Mars is <1 mbar during Step 2. As on today's Mars, the outgassing source is swamped by photochemical sinks during Step 2 (Krasnopolsky et al. 2004). Moreover, what little $CH_{4(g)}$ exists is radiatively ineffective during Step 2. This is because $CH_4$-$CO_2$ collision-induced absorption (CIA) is weak when the collapsed-atmosphere $P_{CO2}$ is low (Wordsworth et al. 2017, Turbet et al. 2019).

Step 3. $CO_2$ atmosphere re-inflation and further $CH_4$ release. Mars atmospheric collapse occurs at low obliquity (Soto et al. 2015). During the interval of atmospheric collapse, $P_{CO2}$ is set by vapor pressure equilibrium with polar temperature. Most $CO_2$ is sequestered as ice (Fig. A2). Obliquity can fall further after collapse is triggered, but obliquity will eventually rise. As obliquity rises, the collapsed atmosphere is still too thin to warm the poles – a hysteresis effect (Fig. A2). Although $H_2O$ ice shields $CO_2$ ice from peak summer temperatures, $H_2O$ ice insulation raises annual-mean polar temperatures, which also increase with obliquity (Fig. 3). When obliquity rises to the point where no value of pressure yields a polar temperature below the condensation point, the $CO_2$ ice caps are no longer stable. As a result, the remaining polar $CO_2$ ice caps rapidly ($10^3$ yr) and completely sublimate (Fig. 6). As a result, the $CO_2$ greenhouse effect rapidly increases planet-wide.

[2] Although the growing polar ($H_2O$+$CO_2$)-ice cap overburden stabilizes polar clathrate, this represents <10% of the planet's surface area and is omitted.

Our scenario assumes that the $CO_2$ in the ice caps is not irreversibly entombed or otherwise irreversibly sequestered. However, if $CO_2$ ice is deposited on top of $H_2O$ ice, then it can be buried by gravitational instability for some parameter choices (Turbet et al. 2017). In our model $H_2O$ ice is deposited on top of $CO_2$ ice, which also acts to shield the $CO_2$ ice. Finally, basal melting can lead to sequestration of $CO_2$ as liquid in the crust. On the other hand, although stiff $H_2O$ ice may enshroud the $CO_2$ ice, flow of thick, soft $CO_2$ ice (Mellon 1996) opens crevasses, which over time will allow $CO_2$ to return to the atmosphere. There is no evidence for large subsurface reservoirs of liquid $CO_2$. Therefore this assumption, while not proven, is reasonable. We also neglect the adsorbed $CO_2$ reservoir in our calculations, because its estimated size (~35 mbar or less; Jakosky 2019) is much less than the ~1 bar needed for a warm climate in our model.

At this point, the low-latitudes have lost some or all of the $H_2O$ ice overburden (much of it migrated away in Step 2). This $H_2O$ ice overburden had the effect of stabilizing $CH_4$ clathrate. $H_2O$ ice returns to the low latitudes slowly: the wait time for all $H_2O$ ice to return to the low latitudes is $>10^6$ yr. The atmosphere fully reinflates in a time much shorter than this: $\sim10^3$ yr. So, without the clathrate-stabilizing effect of $H_2O$ ice overburden, clathrate is now exposed to the $CO_2$ greenhouse warming associated with re-inflation. This $CO_2$ greenhouse warming is strong: for example, ~30 K for a 2 bar atmosphere (Forget et al. 2013, Mischna et al. 2013). As a result, clathrate irreversibly decomposes, releasing $CH_4$. The wait time for $CH_4$ clathrate decomposition is, for $10^2$ m depth to HSZ, equal to (re-inflation time + subsurface conductive-warming time + decomposition time) = ($10^3$ yr + $10^2$ yr + <10 yr) ≈ $10^3$ yr. Therefore, $CH_4$ is outgassed before $H_2O$ ice can re-load the highlands and re-stabilize the $CH_4$ clathrate. The wait time for outgassing is also much less than the $CH_4$ photolysis timescale. Because of rapid release (Fig. A6), the atmospheric concentration of $CH_4$ can be large after Step 3 (e.g., Fig. 6).

The quantity of $CH_4$ release depends on the duration of the atmospheric collapse, the prior unloading history of the $CH_4$ clathrate reservoir, and the volume of the shallow-subsurface $CH_4$ clathrate reservoir. These have the following effect:

- In the collapse-initiated $CH_4$-release scenario, $CH_4$ release is maximal for the first atmospheric collapse in Mars' history that lasts for $\gtrsim$1 Myr (i.e., a collapse so deep that the peaks of individual $10^5$ yr obliquity cycles are not enough to trigger atmospheric re-inflation). This is because such a prolonged collapse gives time for substantial unloading of the $CH_4$-stabilizing water ice overburden from the highlands. For a prolonged collapse, much of the remaining $CH_4$-clathrate reservoir is destabilized upon re-inflation. Inspection of an ensemble of long-term obliquity integrations (Kite et al. 2015), combined with our GCM-derived collapse thresholds (Fig. 3), suggests that the first prolonged collapse happens geologically soon ($\ll 10^9$ yr) after Mars' first-ever atmospheric collapse. By contrast, a large $CH_4$-burst does not result from $<10^5$ yr collapses. For such brief collapses, unloading is so incomplete (according to the low sublimation rates computed using our general circulation model) that most of the highlands clathrate is still stabilized by the highlands $H_2O$ ice that is still in-place at the time of re-inflation.

- In our model, once degassed, pore space is not recharged with $CH_4$, because diffusion of $CH_4$ through cold clathrate is slow. Thus, build-up of mbar of $CH_4$ in the atmosphere by the collapse-initiated $CH_4$ release mechanism is only plausible during the first $\ll 1$ Gyr of Mars' atmospheric-collapse history. (This point presumes that $CH_4$ comes from deep hydrothermal alteration of ultramafic rock. However, if pore space is filled with some amount of liquid $H_2O$, olivine-hosted fluid inclusions can form abiotic $CH_4$ from $H_2$ produced via fluid-inclusion-scale serpentinization as well as radiolysis over relatively short (sub-Gyr) timescales.

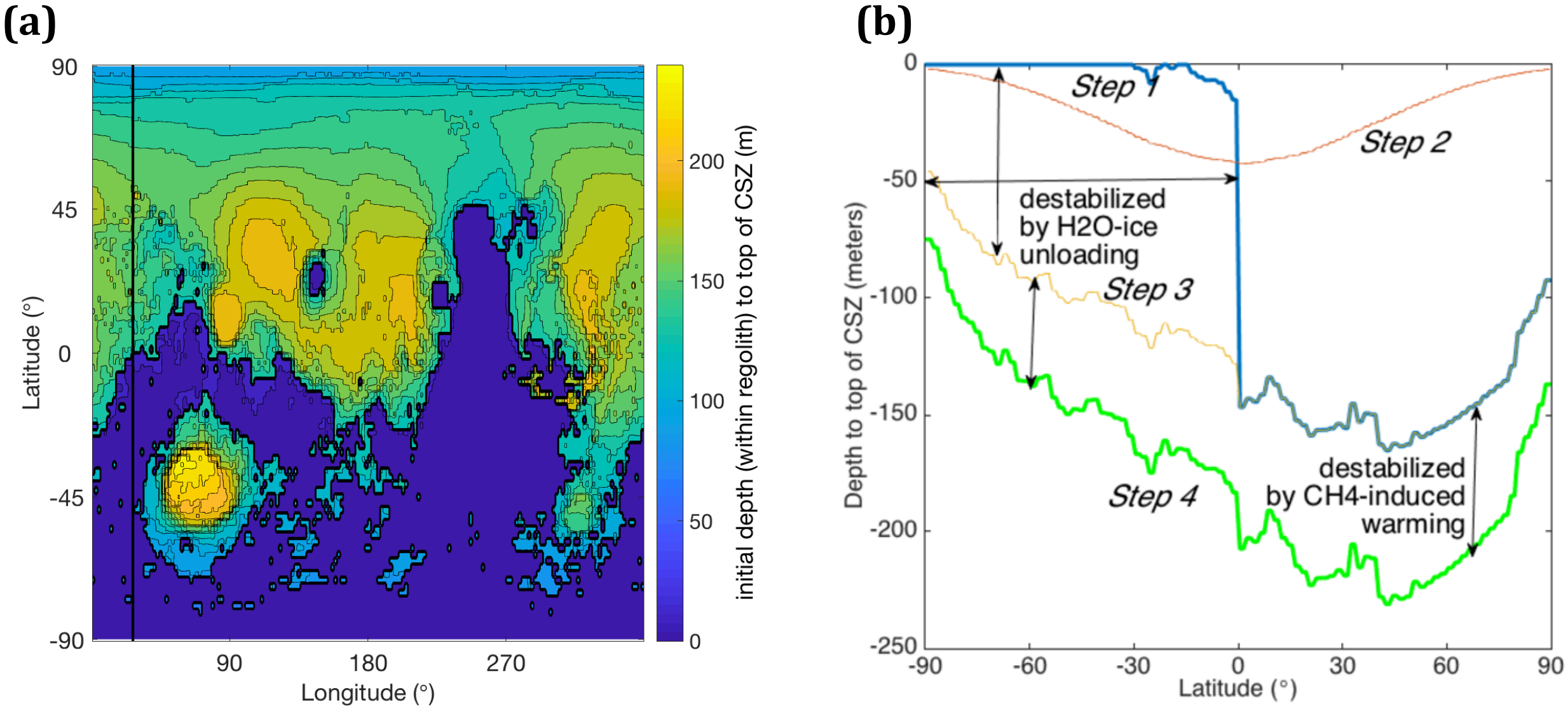

**Fig. 5. (a)** Map of initial depth (within regolith), in meters, to the top of the $CH_4$-clathrate hydrate stability zone (HSZ). This depth is zero where $H_2O$ ice (located at topographic elevations >+1 km, following Fastook & Head 2015) is thick enough to stabilize $CH_4$ clathrate throughout the regolith. Vertical line at 27°E is shown in cross-section (b). **(b)** Latitude-depth cross-section showing the depth to the top of the $CH_4$-clathrate hydrate stability zone. Blue line corresponds to initial HSZ boundary. Red line corresponds to cold, post-collapse conditions, assuming complete $H_2O$ movement to the poles. Yellow line corresponds to re-warmed, post-reinflation conditions, assuming $H_2O$ has not yet returned from the poles, but not including $CH_4$ warming. Green line shows effect of 10 K of further warming (e.g., due to $CH_4$-$CO_2$ CIA).

- The initial (pre-collapse) fraction of the hydrate stability zone that is occupied by $CH_4$ clathrate is unknown. We define *f* as the fraction of the not-yet-degassed portion of the HSZ that is initially occupied by clathrate. $f/\theta$, where $\theta$ is porosity, gives the volume of pore space that is occupied by clathrate.[3] For $f$ > 6%, the atmosphere contains 2-10% $CH_4$ at the end of Step 3. This mix can allow $CH_4$-$CO_2$ collision induced absorption (CIA) to provide climate warming. In order for CIA-warming to exceed the surface cooling caused by $CH_4$ absorption of sunlight, it is a requirement that the atmosphere has >(0.5-1.0) bar $CO_2$ (Turbet et al. 2019). Throughout this paper, we use the $CH_4$-$CO_2$ Collision Induced Absorption warming appropriate to the experimental measurements of Turbet et al. (2019). We do this because the warming calculations of Wordsworth et al. (2017) are based on calculations, not experiments. If we instead adopt the warming calculated by Wordsworth et al. (2017), then we would predict warming that is about twice as strong (Fig. 6, Fig. A3). Therefore, our choice is conservative, in that it minimizes the effects of $CH_4$ bursts and the positive feedbacks of $CH_4$ outgassing on further $CH_4$ release.

[3] The value of *f* can also be interpreted as a measure of degassing efficiency if some $CH_{4(g)}$ is trapped by permafrost.

Step 4. Possible warming due to $CH_4$-enhanced greenhouse effect? The cocktail of circumstances enabled by Mars' first prolonged atmospheric collapse (a massive $CH_4$ pulse, a thick $CO_2$ atmosphere, low-latitude $H_2O$ ice) can potentially warm Mars (Fig. 6, Fig. A3), but special circumstances are required. $H_2O$ snow falling on the equator (plus any $H_2O$ ice that did not have time to sublimate) encounters high insolation (due to lower latitude) and potentially increases greenhouse forcing. The >255 K mean annual temperature threshold, highlighted by the thick green lines in Fig. 6, is important because it is approximately the lower limit for seasonal meltwater runoff to form valleys that drain into perennial ice-covered lakes, as in modern Antarctica; e.g., McKay et al. 1985, McKay et al. 2005). In Fig. 6 the annual mean temperatures are shown for the ±40° latitude, -2 to +3 km elevation zone (valley network zone; Hynek et al. 2010), which is warmer than the global average.

However, $CH_4$-induced warming will only occur if atmospheric collapse is initiated at >(0.5 – 1.0) bar $pCO_2$. Currently available models of atmospheric collapse (Appendix A) do not show atmospheric collapse at such high values of $pCO_2$ (for Early Mars luminosity). On the other hand, collapse occurs at higher $CO_2$ partial pressure if the surface ice/snow cover is extensive, raising significantly the planetary albedo of Early Mars (Ramirez 2017). Moreover, other GCMs return different results (Fig. A2), and the threshold pressure for atmospheric collapse is sensitive to parameters (e.g., Kitzmann 2016). Therefore, although the hypothesis of Early Mars warming due to atmospheric-collapse initiated $CH_4$ release is not ruled out by our study, it requires special circumstances.

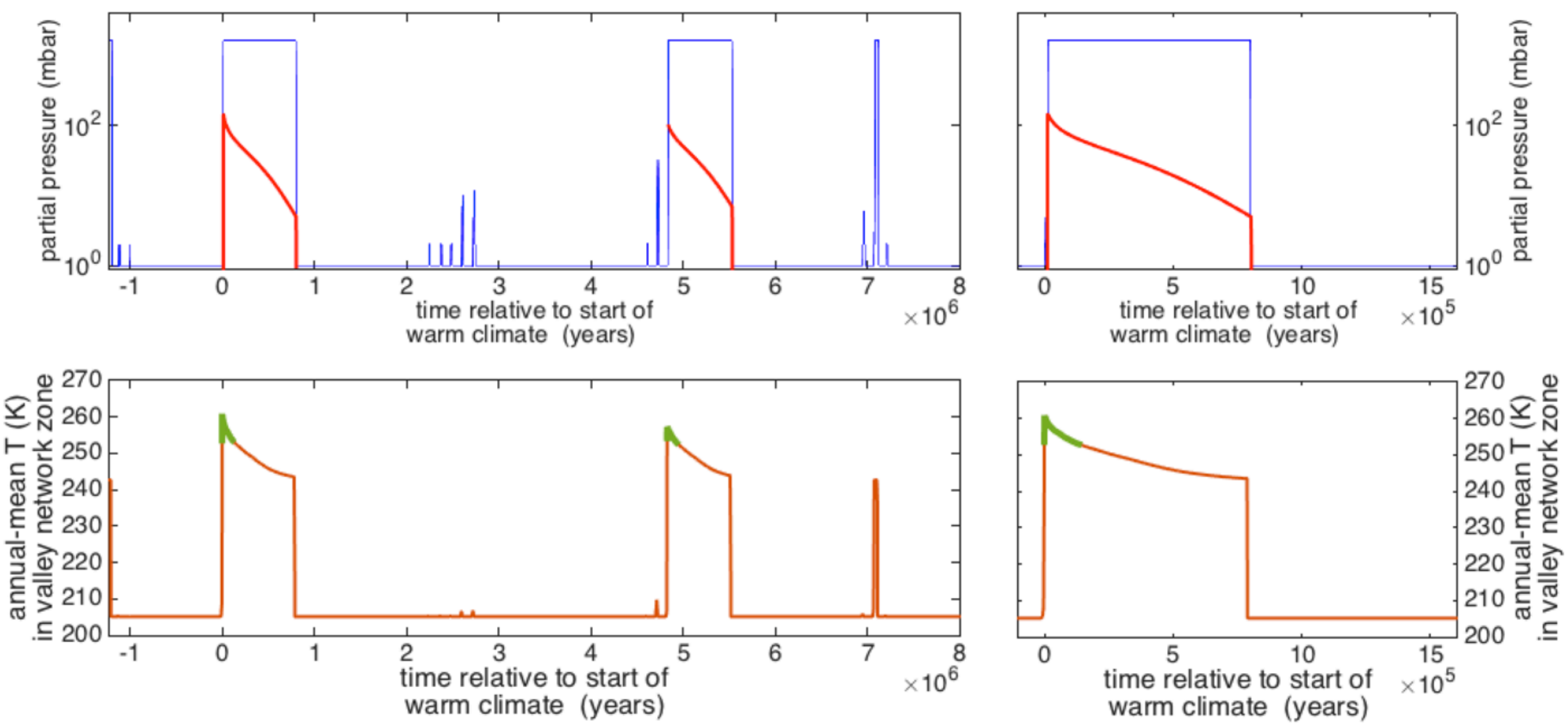

**Fig. 6.** Example climate evolution model output (for *f* = 0.15; sensitivity to *f* is shown in Fig. A4). This is a speculative calculation because it assumes atmospheric collapse is initiated at >0.8 bar $pCO_2$, higher than predicted by currently-available models (Appendix A). *Top:* Pressure evolution. Partial pressure of $CO_2$ (blue) and $CH_4$ (red). *Bottom:* Temperature evolution. After re-inflation, a ~100 kyr-long, >10 K warming occurs. Green highlights the >$10^4$-yr long intervals when annual mean temperatures in the ±40° latitude, -2 to +3 km elevation zone (valley network zone; Hynek et al. 2010) exceed those for lakeshore weather stations in Taylor Valley, Antarctica (Doran et al. 2002). This calculation is for the $CH_4$-induced warming of Turbet et al. (2019); similar calculations using the parameterization of Wordsworth et al. (2017) lead to longer, more intense warm periods.

We did photochemical calculations to find the lifetime of quickly-released $CH_4$ (Appendix A). We found that $CH_4$ can remain in the atmosphere at radiatively important levels for $10^5$-$10^6$ yr. If $CH_4$ permits warming, then this is long enough for frozen ground under lakes to unfreeze, linking lake water to deep aquifers. Such connections might help to reconcile a generally cold Early Mars climate with evidence for groundwater flow through fractures within sedimentary rocks on Mars (e.g., Yen et al. 2017, Frydenvang et al. 2017, Grasby et al. 2014, Ward & Pollard 2018).

The end of the methane pulse: $CH_4$ destruction ($10^5$-$10^6$ yr). Photochemical processes destroy atmospheric $CH_4$ (Fig. A7). Eventually, the atmosphere will again collapse (Fig. 6). The newly collapsed atmosphere can retain many mbar of $CH_4$. That is because at the temperature where nearly all the $CO_2$ has condensed, none of the atmospheric $CH_4$ has condensed. As a result, the atmosphere $CH_4/CO_2$ ratio can transiently exceed the threshold for Titan-like photochemistry, haze, and soot (McKay et al. 1991, Haqq-Misra et al. 2008). This sooty phase is brief, because $CH_4$ destruction is rapid when $CH_4/CO_2$ is high (Fig. A7). If all $CH_4$ is converted to soot while $CH_4/CO_2 > 0.1$, then the soot layer column mass can be >100 kg/m$^2$. This is sufficient to contaminate the crust at the organic matter levels reported by Eigenbrode et al. (2018) to a depth of ~7 km, or to 300 m depth for the ~0.1 wt% levels suggested by analysis of the temperature spectrum of the SAM EGA (Sutter et al. 2017).

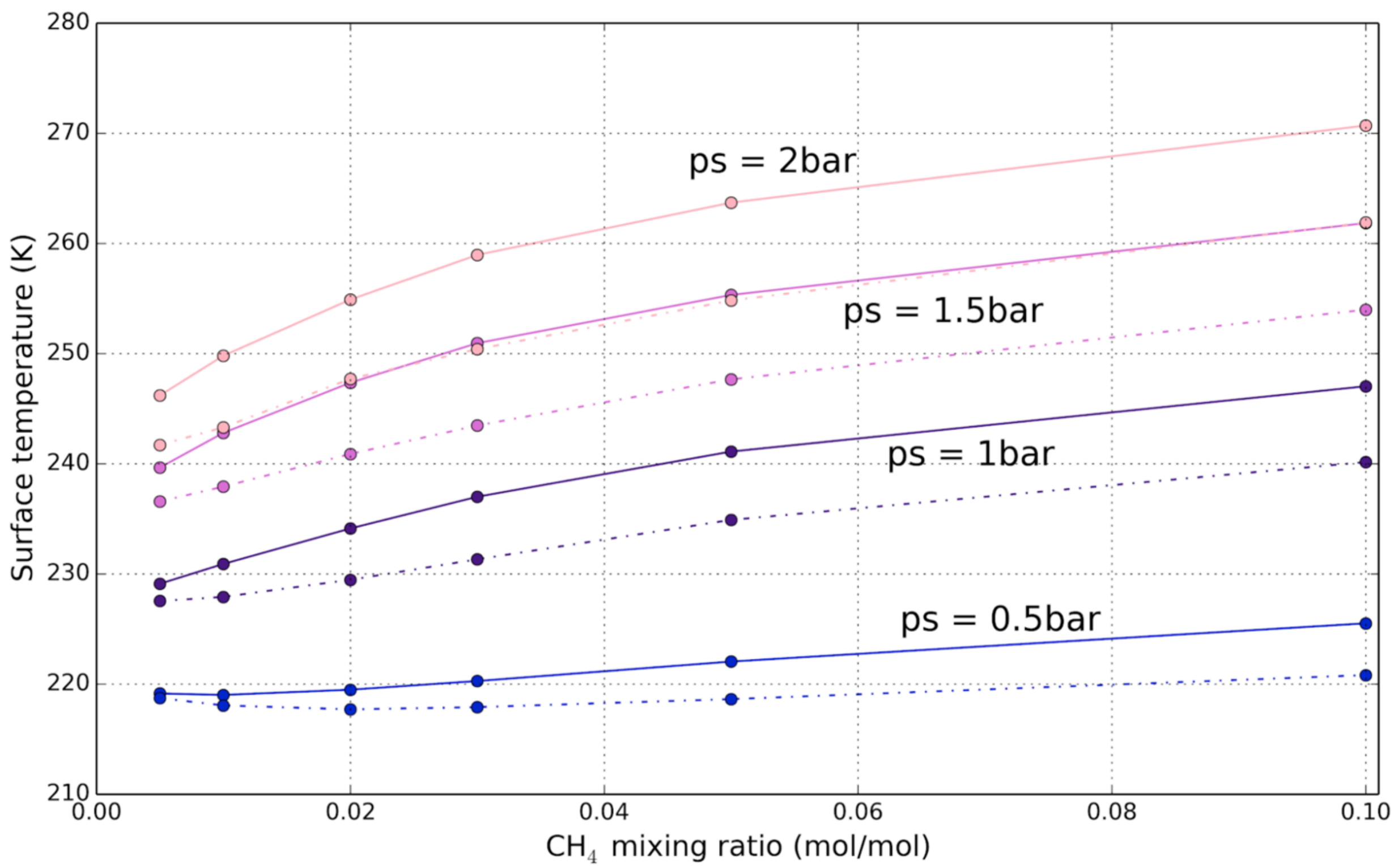

**Fig. 7.** $CH_4$ warming plot. ps = partial pressure of atmospheric $CO_2$ at Mars' surface. Dashed lines correspond to warming calculated on the basis of experiments by Turbet et al. (2019), and solid lines correspond to warming calculated on the basis of calculations by Wordsworth et al. (2017). If the Wordsworth et al. (2017) results are used, then stronger $CH_4$-induced climate warming would be obtained. Note that for ps = 0.5 bar, adding $CH_4$ can cause cooling. This is due to absorption of near-infrared insolation by $CH_4$.

## 4. Discussion.

### 4.1. Sensitivity tests show that special conditions are needed for collapse-initiated $CH_4$ release to cause warming.

One key control for determining whether atmospheric-collapse-initiated $CH_4$ release can cause surface warming is the fraction of HSZ volume occupied by clathrate, $f$ (Fig. 8, Fig. A4). For $f < 0.01$, strong surface warming is precluded. For $f > 0.1$, warming can be so strong that snowmelt can feed lakes. Methane is released by the warming triggered by prior $CH_4$ release. This feedback $CH_4$ release increases with *f*. The positive feedback on $CH_4$-induced warming can be large when *f* > 0.1 (Fig. 8).

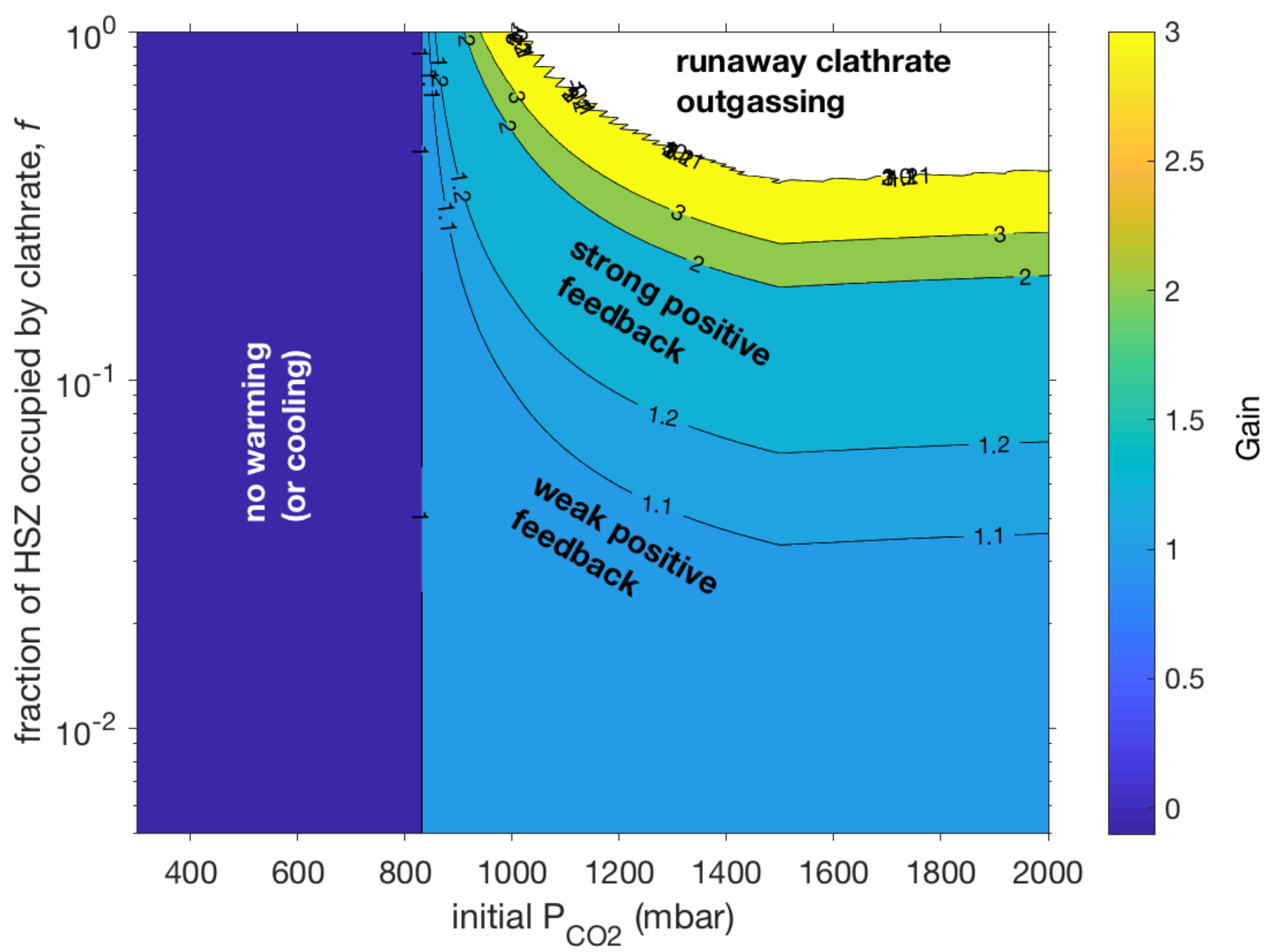

**Fig. 8.** Map of feedback strength (color shading corresponds to gain as defined in Eqn. 1) for a single-column model, assuming initial surface temperature of 240 K. Values of *f* > 0.3 exceed plausible porosity and so are unlikely.

For the calculations underlying Fig. 8, we find the runaway clathrate breakdown threshold by comparing direct $CH_4$-induced warming to the feedback warming (induced by the $CH_4$ release driven by direct warming). If the indirect, feedback warming exceeds the direct warming, then the system will run away (i.e., all $CH_4$ clathrate will be destroyed). This only occurs for *f* > 0.3, which exceeds plausible porosity and so is unlikely. Otherwise the asymptotic warming gain (G) is given approximately by

$$G = 1/(1-R) \qquad (1)$$

(linearizing the feedback), where R is the feedback factor (Roe 2009). Fig. A5 shows results including nonlinear feedbacks.

In our GCM, atmospheric collapse can be avoided (even for zero obliquity) down to initial pressures <0.6 bar. This is unfavorable for collapse-initiated $CH_4$-$CO_2$ climate warming. The collapse trigger atmospheric pressure has to be >(0.5-1.0) bar, otherwise surface warming from $CH_4$ is weak or even net-negative (Turbet et al. 2019; Fig. 8, Fig. 7).

$CH_4$ release increases strongly with initial *T*. This is because of the nonlinearity of the $CH_4$ clathrate decomposition curve (Fig. 4, Fig. A5).

The maximum size of a $CH_4$ burst is reduced if the HSZ partially outgasses prior to the main burst. Pre-warm-climate outgassing can occur due to obliquity variations (Kite et al. 2017a), or as the result of ice unloading during pre-main-burst atmospheric collapses. We used an ensemble of long-term obliquity simulations (Kite et al. 2015) to explore these effects. In these calculations, we made the simplifying assumption that $CH_4$ release was proportional to the volume of $H_2O$ ice removed, but with $CH_4$ release suppressed from depths that had previously been degassed. This simplifying assumption allowed us to carry out >30 long-term simulations. We determined the fraction of runs for which at least 100 m of $H_2O$ ice first-time unloading occurs during a single atmospheric collapse interval, and the dependence of this fraction on the obliquity for re-inflation and on the ice sublimation rate. The 100 m first-time ice-unloading value was chosen as a threshold for strong $CH_4$ outgassing. The "first-time" qualifier refers to the high probability that the $CH_4$-clathrate reservoir is not recharged by hydrothermal circulation. Because of the lack of recharge by this mechanism, repeated ice unloading will only yield $CH_4$ gas from a given latitude-longitude-depth volume element for the first time that volume element is unloaded (considering only the hydrothermal-circulation mechanism for $CH_4$ production; other mechanisms are possible, e.g. Klein et al. 2019). We found that the fraction of runs with a strong $CH_4$ burst increases steeply when the difference between the obliquity for collapse and the obliquity for re-inflation increases (i.e. when the "height" of the hysteresis loop on Fig. A2 increases). However, the fraction of runs with a strong $CH_4$ burst is insensitive to the ice unloading rate.

Consistent with the results of GCM simulations (e.g., Soto et al. 2015), the scenario assumes that $CO_2$-atmosphere collapse occurs only for $\phi$<40°. Otherwise, no major $H_2O$ ice shift will occur at collapse, because $H_2O$ ice is stable at low latitudes for low $P_{CO2}$ when $\phi$>40°.

Overall, our conclusion is that special circumstances are required in order to uncork $10^{17}$ kg of $CH_4$, the minimum needed for strong warming.

## 4.2. Predictions.

If atmospheric collapse initiated a methane burst on Early Mars, then this would have consequences for Mars geology. These consequences lead to testable predictions. For example, diameter < 50 m impact craters that predate the methane burst should be very rare or absent (Kite et al. 2014, Warren et al. 2019, Vasavada et al. 1993). This is because if $P_{CO2}$ was high before the methane burst as required to avoid atmospheric collapse, then small impactors would have been screened by the atmosphere, and so small hypervelocity impact craters would not form. In addition, pre-methane-burst large wind ripples (Lapôtre et al. 2016) should be absent, because these only form at low atmospheric density.

Clathrate destabilization on Mars has been proposed to explain both chaos terrain and also mounds interpreted as mud volcanoes (e.g., Milton 1974, Baker et al. 1991, Komatsu et al. 2000, Skinner & Tanaka 2007, Kite et al. 2007, Komatsu et al. 2016, Ivanov et al. 2014, Oehler & Etiope 2017, Pan & Ehlmann 2014, Etiope & Oehler 2019, Brož et al. 2019). Clathrate destabilization has also been proposed as a trigger for the formation of chaos terrain.

If methane bursts occurred on Early Mars, that could also entail new interpretations of Mars geochemistry. One possibility is deposition of >100 kg/m$^2$ of photochemical soots. Such soots can indicate a reducing past atmosphere ($CH_4/CO_2 \gtrsim 0.1$) (Trainer et al. 2006, Haqq-Misra et al. 2008). $CH_4/CO_2 \gtrsim 0.1$ is predicted to occur immediately following atmospheric recollapse. Other mechanisms for rapidly delivering $CH_4$ to the Mars atmosphere, such as large-amplitude changes in mean obliquity (Kite et al. 2017a) or impact delivery (Haberle et al. 2018), might also produce photochemical soots. This prediction is testable using the SHERLOC (Scanning Habitable Environments with Raman & Luminescence for Organics & Chemicals) instrument on the Mars 2020 rover (Beegle et al. 2015). Soots (complex abiotic organic matter) within ancient sediments are a potential life detection false positive (Sutter et al. 2017, Neveu et al. 2018, Eigenbrode et al. 2018).

## 4.3. Implications.

Our collapse-initiated $CH_4$-release scenario, which considers only $CH_4$ produced by serpentinization, only works if serpentinization occurred on Mars (Lasue et al. 2015). However, the fraction of the crust that must undergo serpentinization is small (<0.1%). Indeed, although serpentine is detected on Mars, these detections are overall uncommon (Ehlmann et al. 2010, Amador et al. 2018, Leask et al. 2018). This is not surprising, because serpentinization may be mostly restricted to rocks too deep for subsequent exhumation to the surface (Carter et al. 2013, Sun & Milliken 2015). Moreover, radiolysis+fluid-inclusion-scale serpentinization can produce $H_2$+$CH_4$ and not leave behind a mineralogic byproduct that would be detectable by CRISM (Klein et al. 2019). Regardless of production depth and process, $CH_4$-charged waters may be swept to the near-surface by hydrothermal circulation (Parmentier & Zuber 2007).

The collapse-initiated $CH_4$-release scenario tends to produce, at most, one methane spike. If this occurred >0.5 Gyr after Mars formation, then the warm climate could be associated with valley network formation (Fig. 9). Previously proposed triggers for warm climates on Mars may supplement the mechanism discussed here and are not inconsistent with it (e.g., Kite 2019 and references therein).

In the collapse-initiated $CH_4$ release scenario, a $CH_4$-induced warm climate (if it occurs at all) is preceded and postdated by a cold climate with a thin atmosphere. This has broadly negative implications for Mars surface astrobiology. For astrobiology rovers, fluviolacustrine deposits ascribed to the Noachian/Hesperian boundary represent candidate landing sites with good potential for organic preservation and habitability (Summons et al. 2011, Ehlmann et al. 2008). Indeed, the delta in Jezero crater is the planned landing site for the Mars 2020 rover (Fassett & Head 2005, Ehlmann et al. 2008, Goudge et al. 2018). However, in the collapse-initiated $CH_4$-release scenario, Mars' surface is sterilized (temperatures ~200 K and >1 MRad surface radiation doses; Hassler et al. 2014) just before the lakes are filled. The lakes might nevertheless be inoculated with life from hot-spring or subsurface refugia.

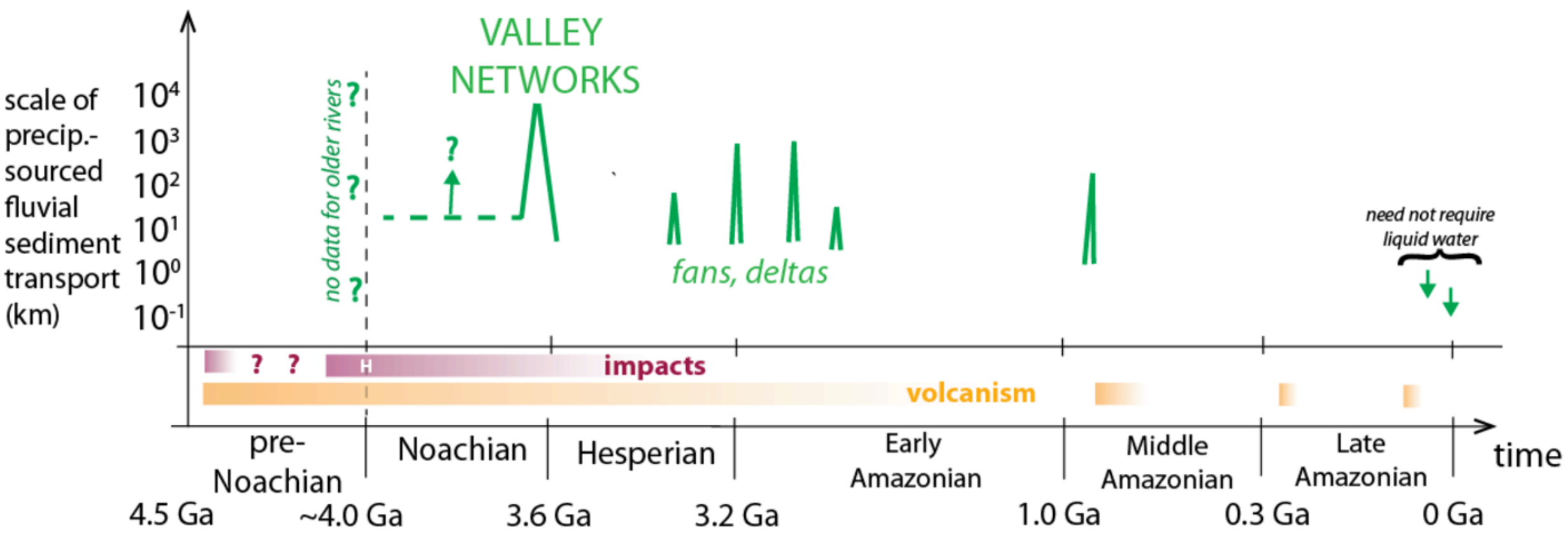

**Fig. 9.** Warm paleoclimates indicated by geomorphology on Mars (modified after Kite et al. 2017b). Estimated ages are from Michael (2013). Y-axis corresponds to the map-view scale of the landforms shown. Neither the durations of geologic eras, nor the durations of river-forming climates, are to scale. Data are consistent with long, globally dry intervals.

## 5. Conclusions.

We investigated a mechanism in which rapid and massive (>30 mbar) degassing of $CH_4$-clathrate occurs consequent to atmospheric collapse and atmospheric reinflation. In this collapse-initiated $CH_4$-release scenario, atmospheric collapse causes low-latitude surface water ice to sublimate away, depressurizing and thus destabilizing $CH_4$ clathrate in subglacial pore space. Subsequent atmospheric re-inflation leads to warming that further destabilizes $CH_4$ clathrate, leading to more outgassing. The $CH_4$ bloom is brought to a close by photochemical destruction of $CH_4$ (potentially accelerated by a new atmospheric collapse). During the $CH_4$ bloom, in some of our model runs, $CH_4/CO_2$ transiently exceeds the threshold for haze and soot – a "Titan-like" blip in Early Mars history. Drawdown of the $CH_4$-clathrate reservoir ensures that the first $CH_4$ spike is also the most intense.

Currently-available models indicate that special circumstances would have been required for $CH_4$-induced warming to occur and attain mean annual temperatures >255 K (consistent with perennial ice-covered lakes; McKay et al. 2005). Specifically, the fraction of the not-yet-degassed portion of the clathrate hydrate stability zone that is occupied by clathrate ($f$) must exceed 0.1, and atmospheric collapse must occur for $P_{CO2}$ > 0.8 bar. Our warming model output is also sensitive to the choice of $CH_4$-$CO_2$ CIA opacities: experimentally derived parameters (Turbet et al. 2019) lead to colder outcomes than the parameters included in HITRAN (Wordsworth et al. 2017, Karman et al. 2019).

## Appendix A. Methods.

### A.1. Overview.

In order to model $CH_4$ release initiated by atmospheric collapse and subsequent atmospheric reinflation, we use simulated obliquity forcing to drive a model of surface temperature evolution (Fig. A1). Temperature is calculated as a function of latitude, longitude, and depth beneath the surface. The model uses a GCM-derived parameterization of atmospheric collapse and re-inflation (Fig. A2). For the surface temperature boundary condition, a GCM-derived look-up table is used. Changes in surface temperature and overburden pressure cause subsurface $CH_4$-clathrate to decompose and outgas. Once outgassed, $CH_4$ can be destroyed by photochemical processes. $CH_4$ can either warm or cool the surface, depending on $P_{CO2}$. Different assumptions about long-term escape-to-space of volatiles can be incorporated by varying initial conditions. Once initialized, the total surface-exchangeable $H_2O$ and $CO_2$ inventories (i.e., sum of the atmosphere + ice caps + shallow ground ice reservoirs) are held constant for the duration of a model run ($10^7$-$10^8$ yr). In effect, we assume that neither escape-to-space of $H_2O$ and $CO_2$, nor geologic sequestration of $H_2O$ and $CO_2$, cause a large fractional change in the $H_2O$ and $CO_2$ inventories during the model run ($10^7$-$10^8$ yr).

The total $CH_4$ outgassed due to Mars atmospheric collapse and re-inflation can be decomposed into:

- Prompt $CH_4$ release from atmospheric collapse (reduction in $P_{CO2}$, combined with planetwide cooling). This is always zero, because the large stabilizing effect of cooling defeats the small destabilizing effect of the loss of the weight of the atmosphere.
- Minor $CH_4$ release from $H_2O$ ice unloading following atmospheric collapse (Step 2 in Fig. 2).
- Major, re-inflation $CH_4$ release: from warming during re-inflation (Steps 3-4 in Fig. 2).
- $CH_4$ release (if any) due to warming feedbacks (Steps 3-4 in Fig. 2).

Direct $CH_4$ release is proportional to the fraction of the not-yet-degassed portion of the HSZ that is initially occupied by clathrate ($f$). Feedback $CH_4$ release increases nonlinearly with $f$.

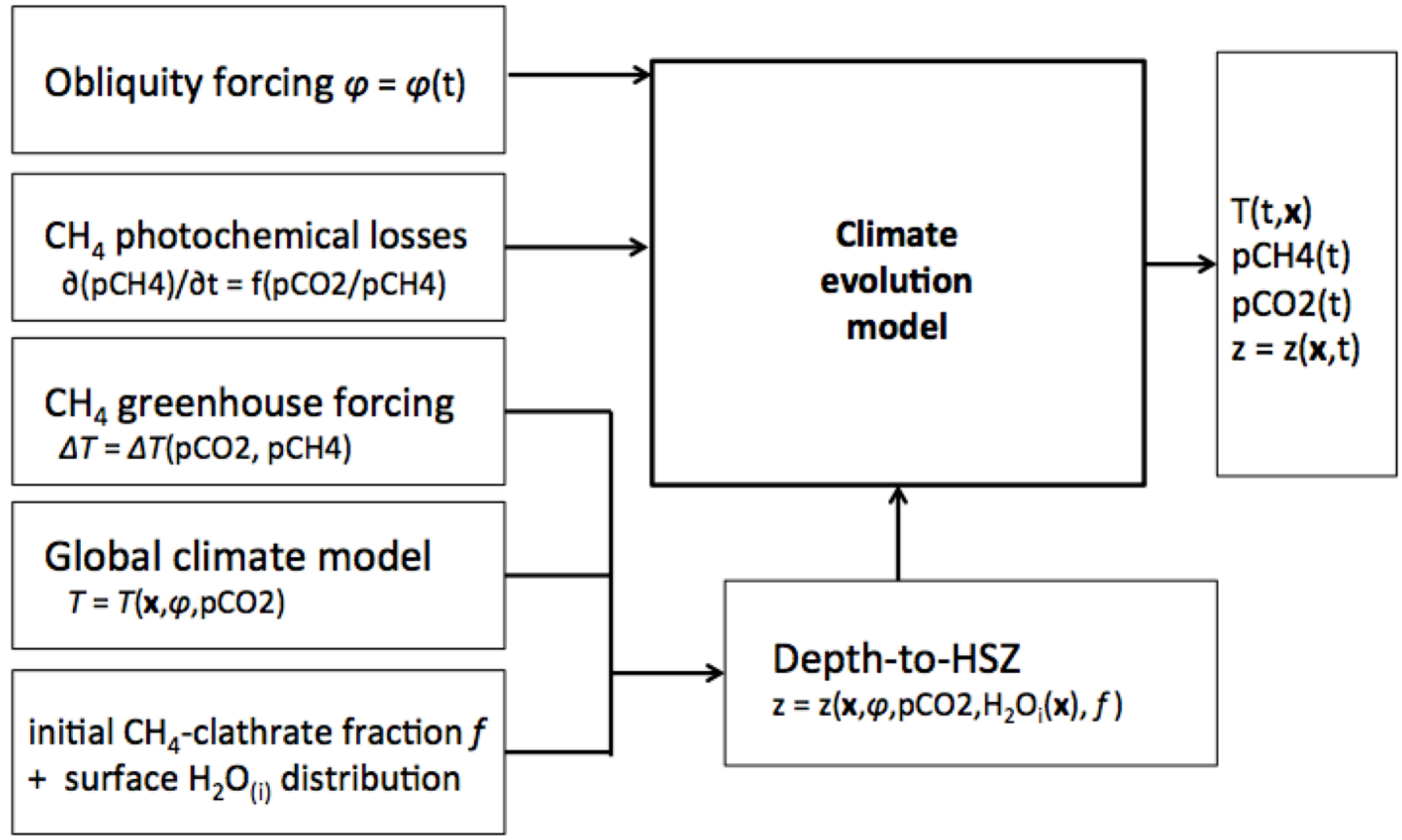

**Fig. A1.** Schematic of the model. HSZ = $CH_4$-clathrate Hydrate Stability Zone

## A.2. Atmospheric Collapse and Surface Temperature Modeling.

To calculate atmospheric collapse as a function of obliquity and $P_{CO2}$ (Fig. A2), we used the Mars Weather Research and Forecasting general circulation model (MarsWRF GCM; Mischna et al. 2013, Richardson et al. 2007, Toigo et al. 2012) to build a look-up table for Mars annual-mean surface temperatures as a function of obliquity, log of $P_{CO2}$, and latitude, for a solar luminosity 75% of modern (corresponding to 0.7 Ga after Mars formation according to the standard solar model; Bahcall et al. 2001). Four $P_{CO2}$ values were investigated: 6 mbar, 60 mbar, 600 mbar, and 1200 mbar. Five obliquities were investigated 0°, 5°, 15°, 25° and 35°. Other values of $P_{CO2}$ and obliquity were obtained by extrapolation and interpolation from this grid of 5 × 4 = 20 runs. In each scenario, we began with a dry atmosphere, and a water ice north polar cap ranging from 1-10,000 kg/m$^2$ thickness depending on the simulation. There is no $CO_2$ ice on the surface to begin with. We assumed a spatially and temporally uniform atmospheric dust opacity of 0.3. This is a modest amount of dust, and ignores the complicating factors of seasonality and strength that are unknowns for different obliquities and surface pressures. For comparison, we also investigated a second scenario with zero dust. All cases contain radiatively active $H_2O$ and $CO_2$ ice clouds where saturation is reached. We used a polynomial fit to these results to determine the boundary of atmospheric collapse (i.e., coldest temperatures at the $CO_2$ condensation point). Radiatively active $CO_2$ ice aerosol is included. $CO_2$ ice grain size radius is assumed to be 100 μm. This choice is based on the argument of Forget et al. (1998) that these larger particles would be appropriate in a $CO_2$-rich atmosphere. Results are likely sensitive to this assumption (Kitzmann 2016). The GCM runs used for this study use a dust scenario that is derived from Mars Global Surveyor data for a year without a planet-encircling dust storm and used in the LMD Mars Climate Database. In the vertical, a Conrath-ν dust profile is used.

Collapse rate is set to ∼1 mbar/yr (Soto et al. 2015). We use the same rate for re-inflation (Soto et al. 2015). We force the model with $\phi$(t) from realistic obliquity histories. We vary $\phi$, but leave the eccentricity and the longitude of perihelion set to present-day values. We do this because $\phi$ is the dominant control on polar insolation (Schorghofer 2008). Collapse takes ∼1 kyr and is followed by a collapsed interval of $10^4$-$10^7$ yr duration and then by ∼1 kyr of re-inflation (Fig. 6). We adopt $\phi_c$ = 16.5°.

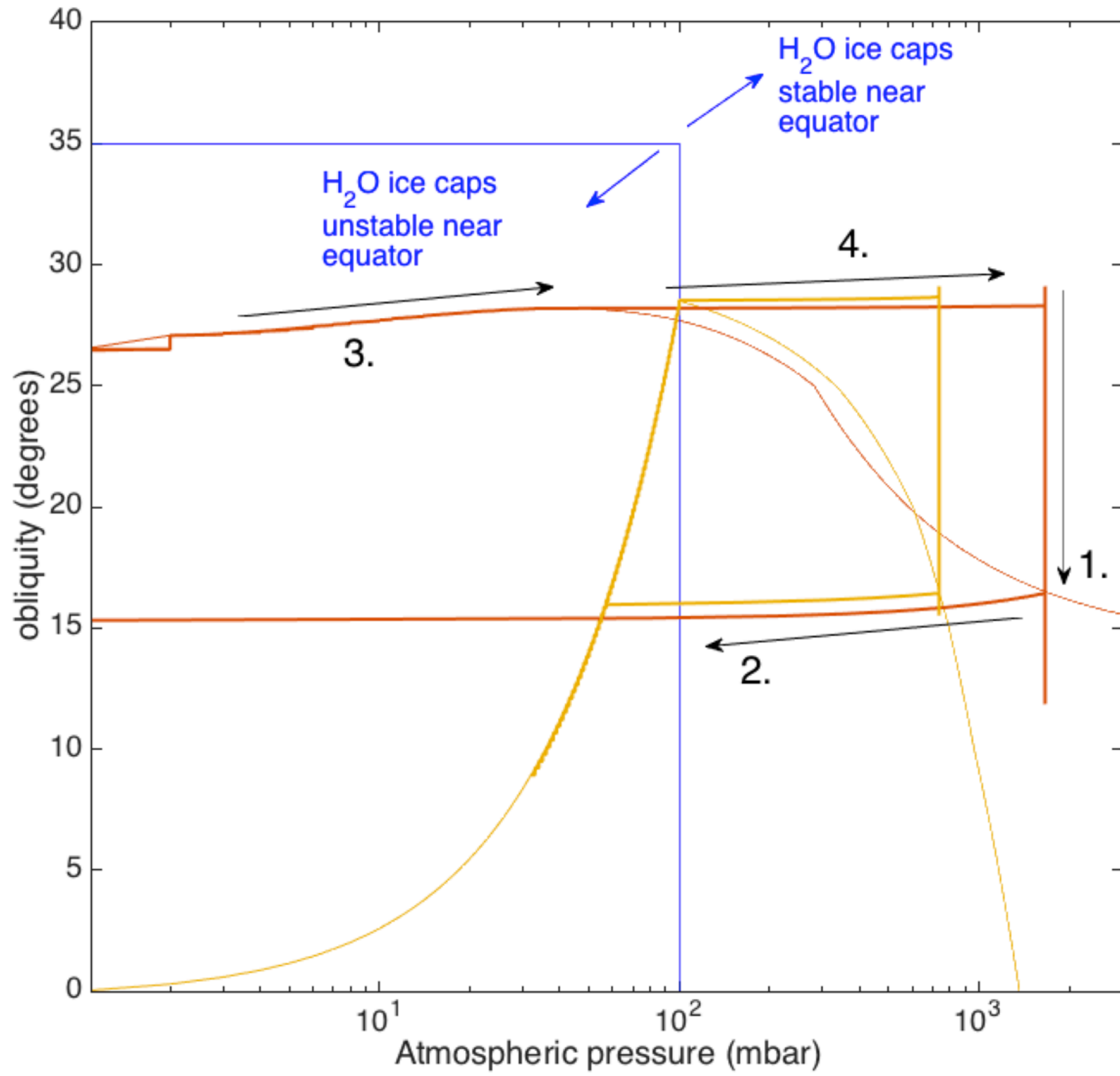

**Fig. A2.** Atmospheric-collapse phase portraits for a collapse-triggering obliquity of ~16.5°, for two different climate models. Results from the GCM of Mischna et al. (2013), with dust, are shown in orange. Loop fit to output of the GCM of Forget et al. (2013) shown in yellow. The hysteresis loops that are highlighted by thick lines are examples of climate system trajectories for a total $CO_2$ inventory of 1.66 bar (orange) and 0.74 bar (yellow) – different conditions are required for the different models. The part of the orange curve that closes the loop from arrow 2 to arrow 3 occurs at very low atmospheric pressure (<0.1 mbar). The thin lines correspond to pressure-temperature combinations on the polar-$CO_2$-condensation curve that are unstable on $10^2$-$10^3$ yr ($CO_2$ condensation) timescales, so are not physically realizable. The high-$P_{CO2}$, low-obliquity vertical "stalk" in the thick orange line corresponds to atmospheres that stay inflated due to $CH_4$-$CO_2$ collision-induced absorption (CIA). These solutions are stable on $10^2$-$10^3$ yr ($CO_2$ condensation) timescales, but decay on $10^5$-$10^6$ yr ($CH_4$ photolysis) timescales.

## A.3. Clathrate modeling and $CH_4$-$CO_2$ warming.

The collapse-initiated $CH_4$-release scenario assumes the existence of shallow $CH_4$-clathrate on Mars, with $CH_4$ produced abiotically in association with deep water-rock reactions (Kite et al. 2017a, Etiope & Sherwood Lollar 2013, Mousis et al. 2013). The $P$-$T$ pathway forming $CH_4$-clathrate is shown in Fig. 4.

Guided by the GCM output, we calculated $T$ and $P$ as a function of depth within the regolith, latitude, and longitude, using $T_{surf}$ from the GCMs, and assumptions about $H_2O$ ice distribution motivated by the water-cycle output of previously published GCMs (e.g., Mischna et al. 2013, Wordsworth et al. 2015). Spatial resolution is (2° latitude) × (2° longitude) × (1 m depth). $T_{surf}$ decreases with elevation, more so as $P_{CO2}$ increases (Wordsworth 2016). The initial $H_2O$ ice distribution is set by allocating $1.4 \times 10^7$ km$^3$ of $H_2O$ ice (Mahaffy et al. 2015) uniformly across land with elevations >1 km (using modern topography). Where $H_2O$ ice rests on top of the regolith, we model the corresponding overpressure and thermal insulation (assuming thermal conductivity, $k_{ice}$ = 2 W m$^{-1}$ K$^{-1}$). The grid is truncated at 300 m depth-within-regolith. This is because destabilization of clathrate to depths >300 m requires annual-mean $T_{surf}$ > 273 K, much warmer than the climate model. For 300 m, the vertical thermal conduction timescale is 1 kyr – much shorter than the other timescales affecting atmospheric $P_{CH4}$. In particular, the $CH_4$ destruction timescale (when the atmosphere is inflated) is $10^5$-$10^6$ yr. This separation of timescales holds even when the low thermal diffusivity of $CH_4$ clathrate and the latent heat of clathrate decomposition are considered. Therefore, we approximate geotherm adjustment as rapid.

Assuming rapid geotherm adjustment, we calculate the depth-to-HSZ. We express this depth as meters below the top of the regolith, so a depth-to-HSZ of 100 m below an ice sheet of 200 m thickness means that clathrate is stable ≥300 m below the surface. We use the phase diagram of Sloan & Koh (2008, their Table 4.1), as shown in Fig. 4. We set porosity = 0.3 (high values of porosity are supported by rover gravimetry; Lewis et al. 2019), assume 120 kg $CH_4$/(m$^3$ clathrate), lithospheric heat flow $Q$ = 0.03 W m$^{-2}$, regolith thermal conductivity $k_{reg}$ = 2.5 W m$^{-1}$ K$^{-1}$, and regolith density $\rho_{reg}$ = 2000 kg m$^{-3}$. For the purpose of modeling the effect of $H_2O$ ice overburden on $CH_4$-clathrate stability, we consider both a case with $H_2O$ ice on high ground and a case without $H_2O$ ice on high ground. The resulting "depth to Hydrate Stability Zone" look-up tables are passed to the main driver.

The main driver uses the following inputs. (1) The depth-to-HSZ look-up tables (Fig. 5). (2) The phase portraits for atmospheric collapse from the GCMs (Fig. A2). (3) A $CH_4$ destruction look-up table (described below) (Fig. A7). (4) A parameterization of greenhouse warming due to $CH_4$ -$CO_2$ collision-induced absorption (CIA) that is based on the experimental results reported in Turbet et al. (2019) (Fig. A4). (5) A simulated obliquity time series obtained using the *N*-body code of Chambers (1999) and an obliquity wrapper script (Armstrong et al. 2004); see Kite et al. (2015) for details. To save computer time, we shift the obliquity time series up and down as needed to trigger collapse relatively early in the run (the likely true wait times for atmospheric collapse depend on the initial obliquity of Mars, and range from only a few Myr if Mars' obliquity was initially low, to hundreds of Myr if Mars' obliquity was initially high).

Fed by these inputs, the main driver first calculates the depth-to-HSZ for four key states (shown in Fig. 5), as follows: (1) pre-collapse state ($H_2O$ ice on highlands, inflated atmosphere – before Step 1 in Fig. 2); (2) immediately post-collapse ($H_2O$ ice on highlands, collapsed atmosphere – end of Step 1 in Fig. 2); (3) fully unloaded but collapsed state (negligible $H_2O$ ice on highlands, collapsed atmosphere – end of Step 2 in Fig. 2); and (4) immediately after re-inflation (negligible $H_2O$ ice on highlands, re-inflated atmosphere – Step 3 in Fig. 2). For some atmospheric collapses, the duration of collapse is too brief for complete unloading of $H_2O$ ice from the highlands. Therefore, we computed compute depth-to-HSZ following re-inflation for a range of time intervals that are too short for complete unloading of $H_2O$ ice. We assume that unloading is steady and uniform (we explored nonuniform unloading specifications, but found only trivial differences). Next, based on a HSZ occupancy fraction, *f*, we compute the volume of $CH_4$ clathrate that decomposes on $CO_2$-atmosphere re-inflation for *complete* $H_2O$-ice unloading of the highlands. Methane decomposition is rapid (Stern et al. 2003, Gainey & Elwood Madden 2012). $CH_4$ clathrate breakdown involves a >14% reduction in solid volume, and we assume fractures allow methane gas released at ≤100 m depth to reach the surface in $\ll 10^5$ yr. The greenhouse warming (if any) corresponding to this $CH_4$ outgassing is calculated (Fig. 7, Turbet et al. 2019). The corresponding maximum $CH_4$ release, including $CH_4$-warming-induced $CH_4$ release, is obtained by iteration. For *incomplete* $H_2O$ ice unloading of the highlands, we linearize both the direct $CH_4$ release in response to re-inflation, and also the $CH_4$-warming-induced feedback $CH_4$ release. Linearization is invalid for the most extreme warming (see Fig. A5), but is reasonable for determining whether or not a warm climate can occur.

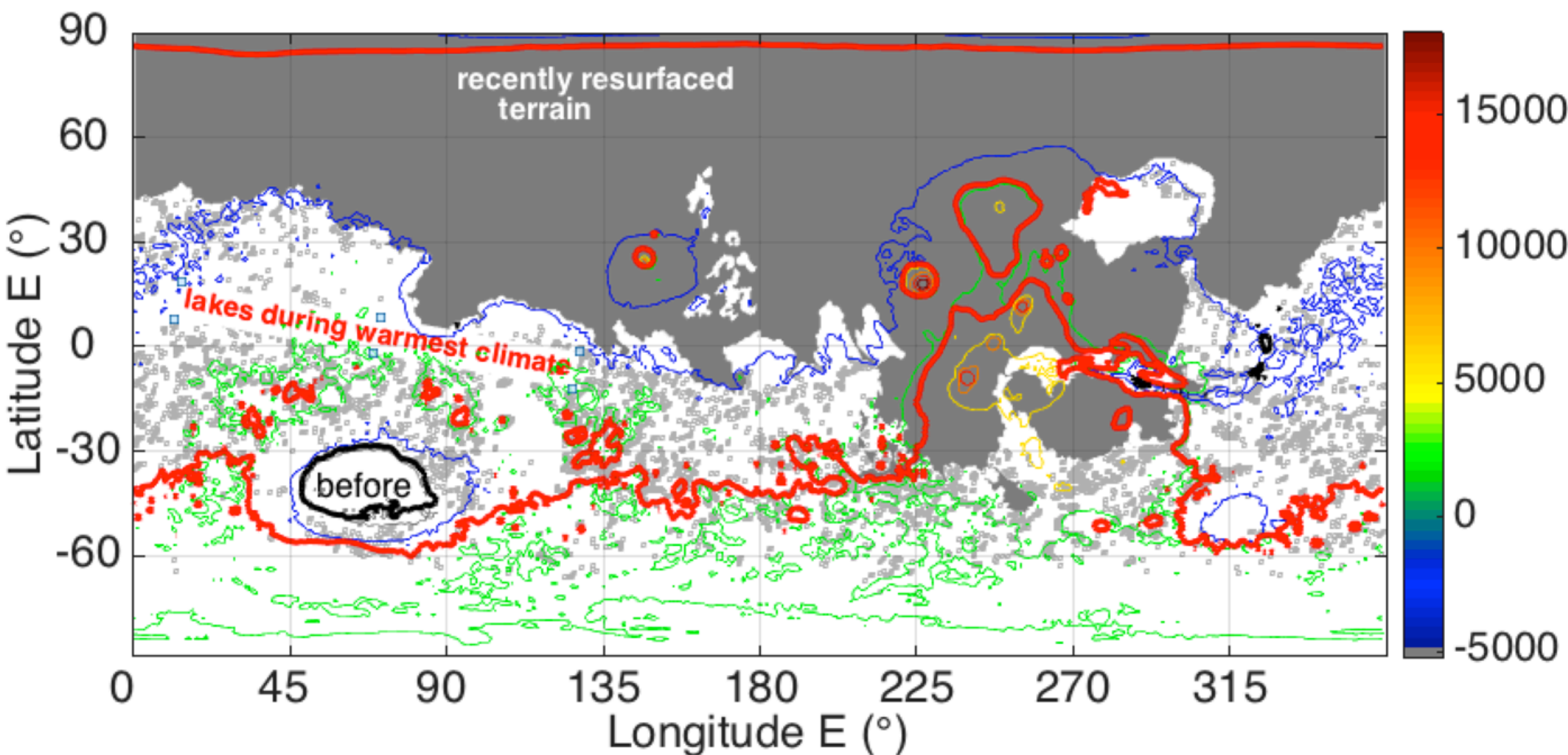

**Fig. A3.** Example warming map showing geographic distribution of rivers and lakes, at peak warming, for the optimistic climate evolution shown in Fig. 6. Special circumstances are required for conditions this warm. Colored contours show topography in meters (present-day topography is used). Before $CH_4$ release, conditions warmer than lakeshore weather stations in Taylor Valley, Antarctica only occur for the lowest elevations (inside thick black contour line), far from mapped valley networks (light gray dots; Hynek et al. 2010). However, with 17 K of uniform $CH_4$-induced warming (Fig. 6), the area warmer than lakeshore weather stations in Taylor Valley Antarctica expands to cover (thick red line) most of the planet (45°S-85°N). Dark gray area is obscured by young lavas.

For large *f*, clathrate decomposition and outgassing can runaway (Fig. A5). Runaway outgassing can produce a Mars climate (Fig. A5) with $T_{ave} > 273$ K – a temperate climate (Halevy et al. 2011, Bishop et al. 2018). A temperate Mars climate lasting ~1 Myr might explain the surface leaching event apparently recorded by aluminous clays overlying Fe/Mg clays at many globally-distributed locations (Carter et al. 2015, Bishop et al. 2018). Runaway clathrate breakdown depends on *f*, the pre-release temperature, and $P_{CO2}$. For cooler pre-release temperatures, the minimum *f* for runaway clathrate breakdown increases, due to nonlinearity in the clathrate phase diagram (Fig. A5).

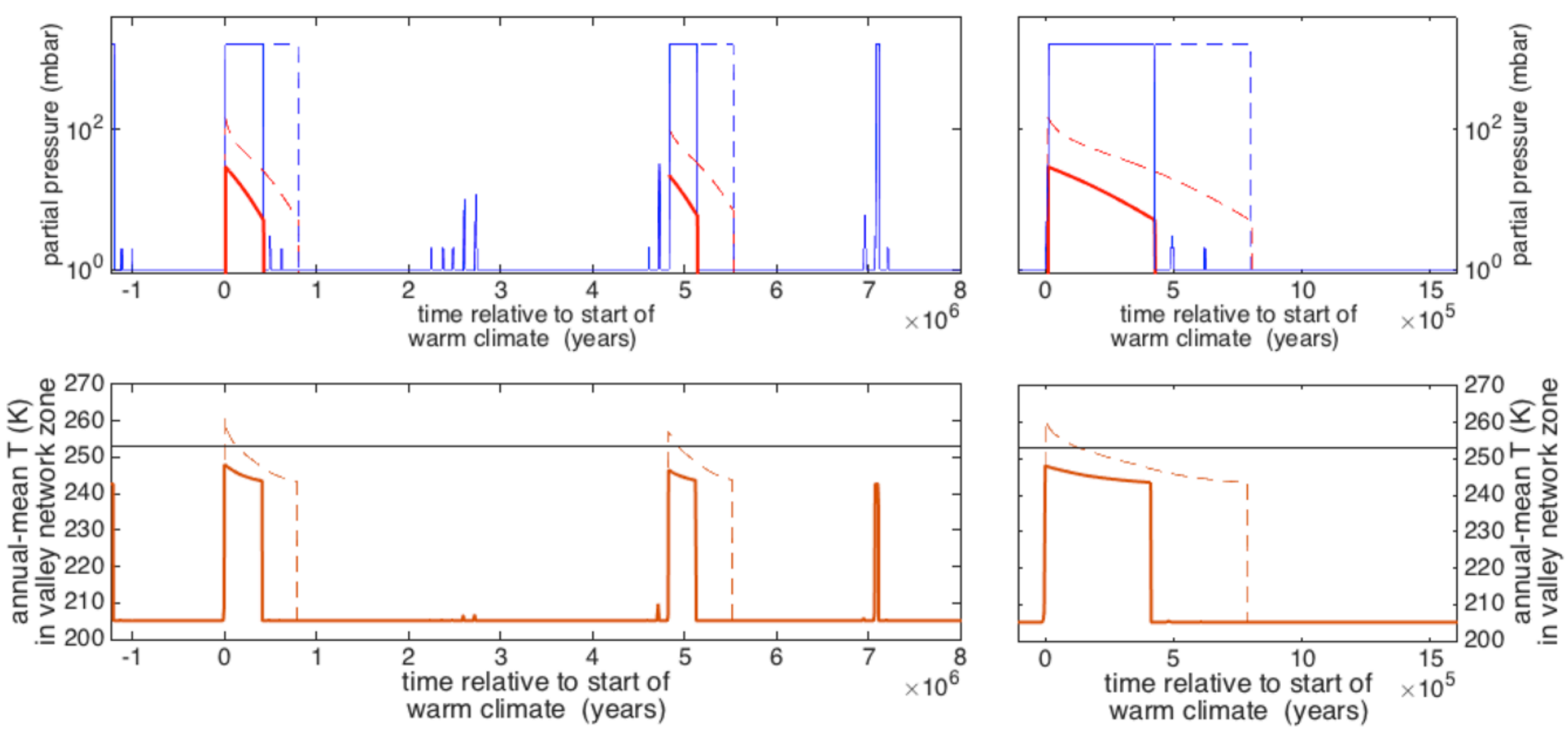

**Fig. A4.** Climate evolution model output. As Fig. 6, but showing sensitivity to a reduction in *f*. Solid lines are for *f* = 0.06, dashed lines are for the value (*f* = 0.15) used in Fig. 6. For *f* = 0.06, the $CH_4$-induced warm climate lasts about half as long, and the peak $CH_4$-boosted annual-average temperatures in the valley network zone do not exceed 253K (horizontal black line in lower panel). Because of these lower peak temperatures, collapse triggered methane outgassing for *f* = 0.06 is unlikely to explain Mars' valley networks.

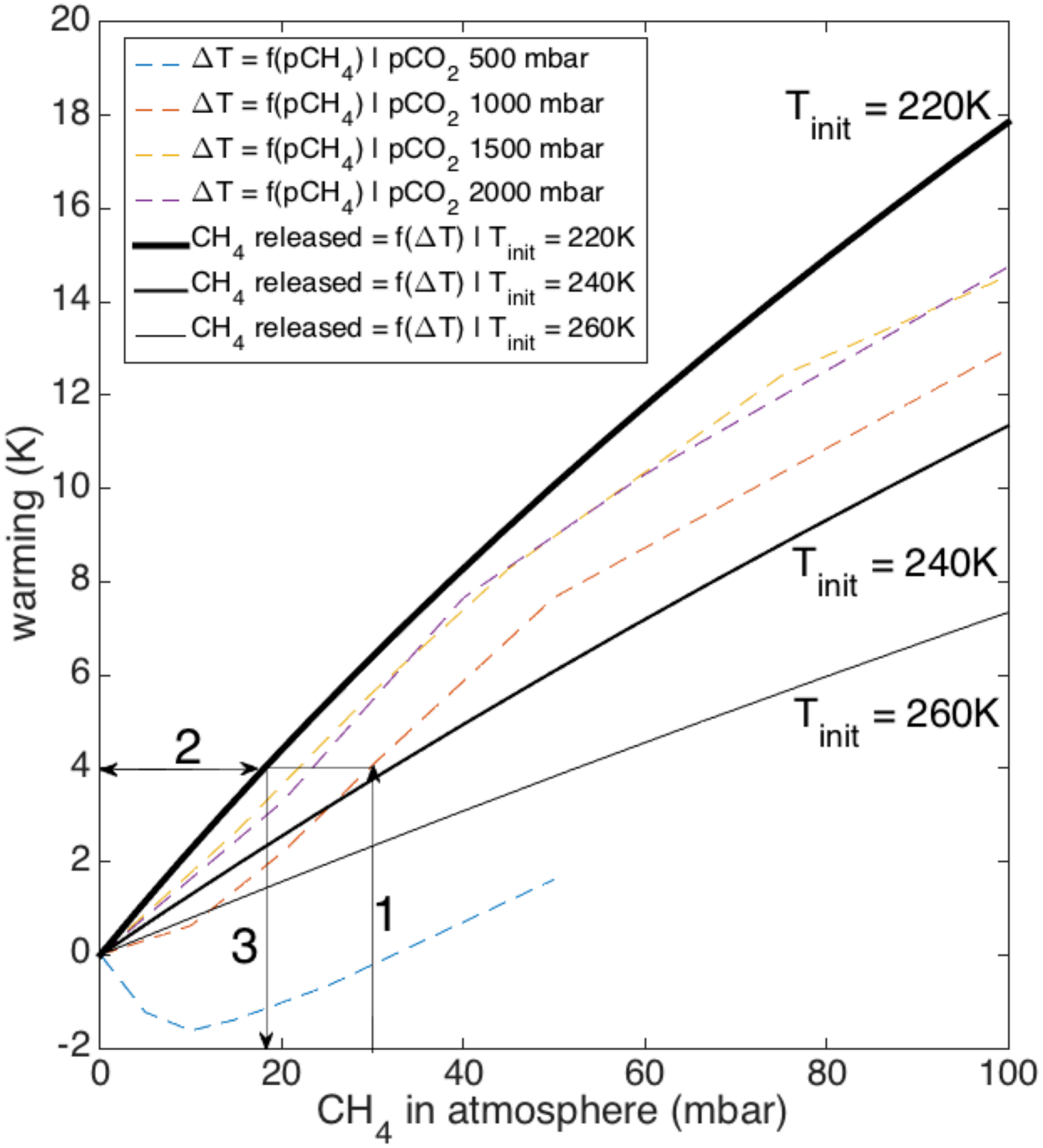

**Fig. A5.** Showing the (narrow) range of conditions under which $CH_4$ outgassing on Mars can lead to runaway warming and outgassing from an $f$ = 0.15 clathrate reservoir. Colored dashed lines correspond to $CH_4$-induced warming (Turbet et al. 2019). Solid lines show the corresponding $CH_4$ release. The arrows labeled 1-3 give an example of how to read the diagram. Supposing (**1**) an initial collapsed initiated $CH_4$ release of 30 mbar $CH_4$ in a 1000 mbar $P_{CO2}$ atmosphere, the warming (**2**) is ~4K. For an initial surface temperature $T_{init}$ = 220K, this is sufficient to release (**3**) a further 18 mbar of $CH_4$, which will lead to further warming and thus more $CH_4$ release. When the warming (dashed line) for a given $P_{CO2}$ plots above a $CH_4$ release line (solid line), runaway outgassing occurs. When the warming for a given $P_{CO2}$ plots below a $CH_4$ release line, there is no runaway, but positive feedback still occurs. These results are for a simplified model using a single column of clathrate-charged regolith.

The time-stepping loop in the main driver uses the phase portraits in Fig. A2, as well as warming due to $CH_4$ (if any), to track atmospheric collapse and re-inflation. Water ice net sublimation rate, $i$, is set to $i$ = 0.15 mm/yr for migration from high ground to poles. This low value (based on MarsWRF GCM results) is due to the low vapor pressure of $H_2O$ in the ~200 K collapsed atmosphere. Movement of ice from poles back to high ground under the warm re-inflated atmosphere should be faster; we use $i$ = -1.5 mm/yr. Our results are qualitatively unaffected by reasonable changes in $i$. The overpressure due to polar ice caps is not explicitly modeled. This is acceptable because polar cap area is small, and, as a result, the fraction of clathrate protected by the polar cap overburden is, likewise, small. For collapse duration $\Delta t \geq (\sim 300\text{ m})/i$, the $H_2O$ ice is completely removed from the highlands. If $\Delta t < (\sim 300\text{ m})/i$, we use a look-up table to find the partial $CH_4$ release. We also track release of $CH_4$ from previous re-inflations and collapses to ensure that the same portion of the regolith cannot release $CH_4$ twice (Fig. A6). Finally, we track inhibition of orbitally paced atmospheric collapses by $CH_4$ warming, if any warming occurs (Fig. 6).

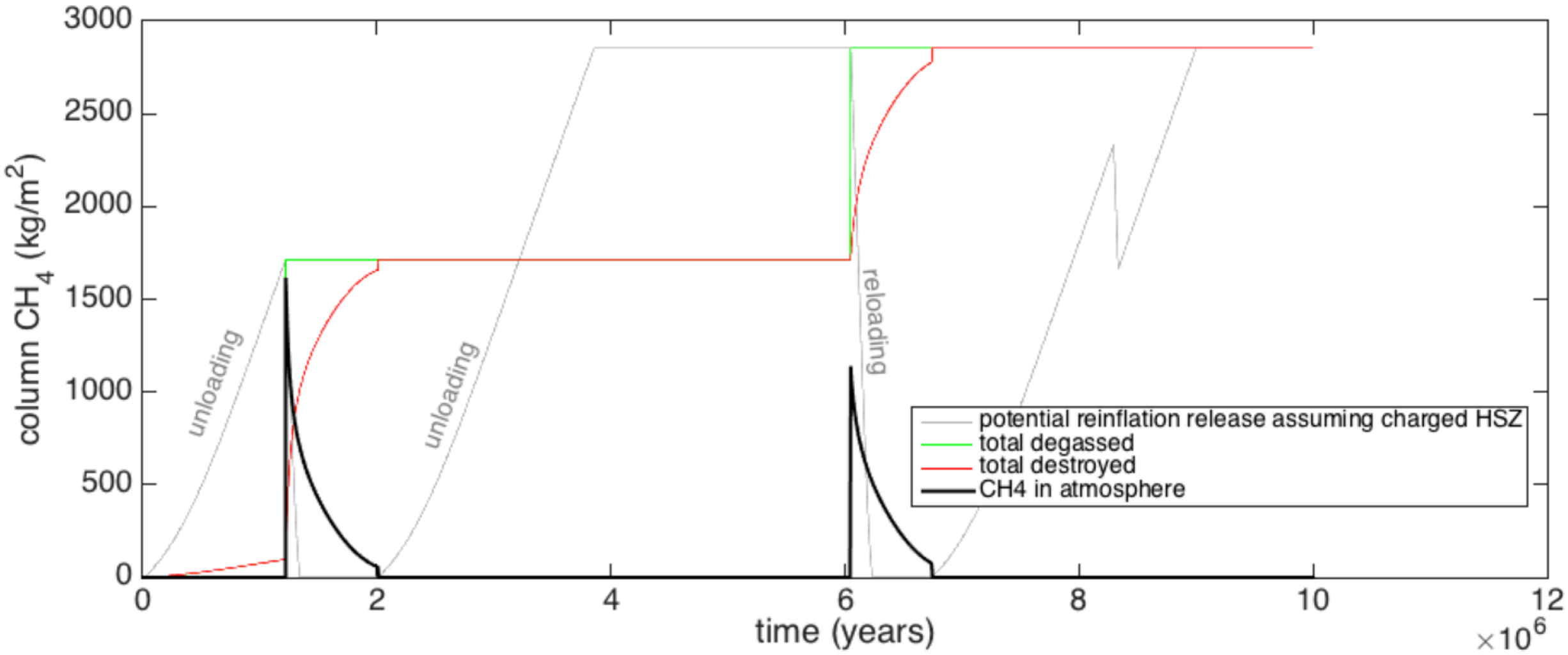

**Fig. A6.** Methane inventory history for one run with $f$ = 0.15. We track release of $CH_4$ from previous re-inflations and collapses to ensure that the same portion of the regolith cannot release $CH_4$ twice.

Relative to climate models that include $CO_2$ and $H_2O$ alone, the amount of warming needed to bring Early Mars valley network locations to temperatures sufficiently high for rivers and lakes is 10 K - 20 K. For warming exceeding 10K-20K, temperatures over most of the Mars Southern Highlands exceed measured temperatures in Taylor Valley Antarctica on the shores of ice covered perennial lakes (Fig. A3) (Doran et al. 2002). Lakes overspill to form regionally integrated valley networks. In Antarctica, a melt-fed lake-overspill river incises into bedrock (Shaw & Healy 1980). High lake-bottom temperatures destabilize sub-lake $CH_4$ clathrates (Fig. 5). Another possible positive feedback, not included in our model, is $H_2$ production by aqueous weathering of olivine at Mars' surface during warm climates (Tosca et al. 2018). $H_2$ is a strong greenhouse gas via $H_2$-$CO_2$ CIA (Ramirez et al. 2014, Ramirez 2017, Turbet et al. 2019).

## A.4. Methane destruction parameterisation.

We used the Caltech/JPL 1-D Mars photochemistry code, modified to include $CH_4$ (and $C_2H_6$, and $C_2H_2$, etc.) (Summer et al. 2002, Nair et al. 1995, Nair et al. 2005), to build a look-up table for $CH_4$ destruction by photochemical processes. The $CH_4$ destruction model is the same as that in Kite et al. (2017a), except for the use of a UV flux appropriate for the Sun 3.8 Ga (Claire et al. 2012). Boundary conditions include surface burial of $O_2$, $O_3$, $H_2O_2$, and CO. Water vapor is set to saturation at the surface; results are insensitive to $H_2O$ concentration. As in Kite et al. (2017), we found that the main control on $CH_4$ destruction rate is the $CH_4$/$CO_2$ ratio, so we fit a curve to the $CH_4$ destruction rate as a function of the $CH_4$/$CO_2$ ratio for use in the climate-evolution model (Fig. A7).

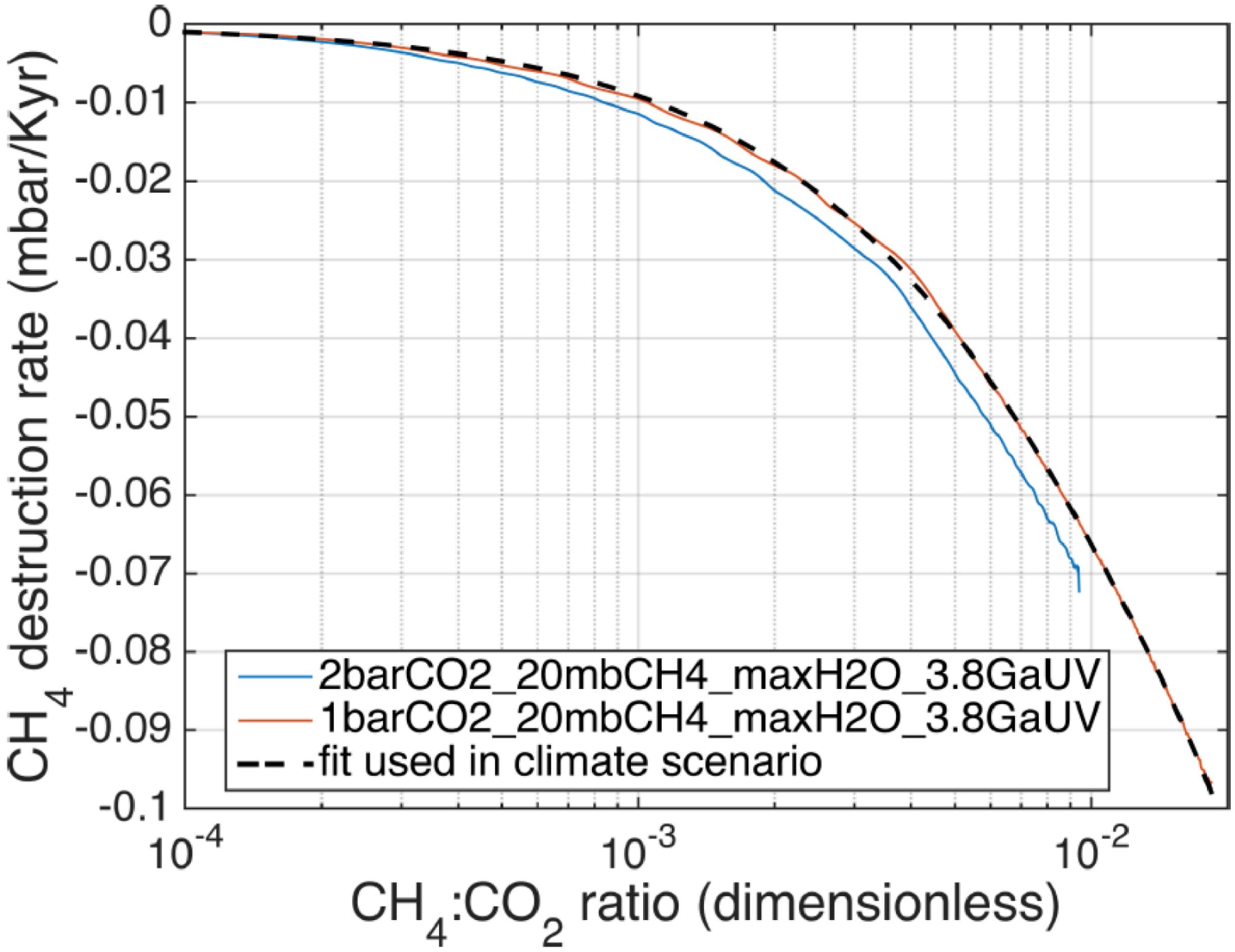

**Fig. A7.** $CH_4$ destruction rate. Both calculations adopt a UV flux appropriate for the Sun 3.8 Ga (Claire et al. 2012), and a high $H_2O$ vapor mixing ratio. Because a high $H_2O$ vapor mixing ratio tends to speed the destruction of $CH_4$ in our photochemical calculations, our use of a high $H_2O$ vapor mixing ratio is conservative in that it will tend to understate the duration of $CH_4$-induced warming.

## Acknowledgements.

We thank Jesse Tarnas, and an anonymous reviewer, for helpful reviews that led to an improved manuscript. We thank Alan Howard, Colin Goldblatt, Ross Irwin, Bob Craddock, Alejandro Soto, John Armstrong, Feng Tian, Itay Halevy, Alex Pavlov, Tom McCollom, Sarah Stewart, and Chris Oze for discussions, and Robin Wordsworth for sharing model output. The MATLAB scripts and GCM summary output used to make the figures in this paper may be obtained for unrestricted further use by contacting the lead author. This project has received funding from the European Union's Horizon 2020 research and innovation program under the Marie Sklodowska-Curie Grant Agreement No. 832738/ESCAPE. M.T. and E.S.K. acknowledge support from the France And Chicago Collaborating in The Sciences (FACCTS) program. E.S.K. acknowledges funding from NASA (NNX16AG55G).

## References.

Aharonson, O.; Zuber, M.T.; Rothman, D.H.; Schorghofer, N.; Whipple, K.X. (2002) Drainage basins and channel incision on Mars, Proc. Natl. Acad. Sci. 99, 1780-1783.

Amador, E.S.; Bandfield, J.L.; Thomas, N.H., 2018, A search for minerals associated with serpentinization across Mars using CRISM spectral data, Icarus, Volume 311, p. 113-134.

Andreassen, K., et al. (2017), Massive blow-out craters formed by hydrate-controlled methane expulsion from the Arctic seafloor, Science, 356, 6341, 948-953, DOI: 10.1126/science.aal4500.

Armstrong, J.C., Leovy, C.B., Quinn, T. (2004), A 1 Gyr climate model for Mars: new orbital statistics and the importance of seasonally resolved polar processes. Icarus 171, 255-271.

Bahcall, J.N., Pinsonneault, M.H., Basu, S., 2001. Solar models: Current epoch and time dependences, neutrinos, and helioseismological properties. Astrophys. J. 555, 990–1012.

Baker, V. R.; Strom, R. G.; Gulick, V. C.; Kargel, J. S.; Komatsu, G. (1991), Ancient oceans, ice sheets and the hydrological cycle on Mars, Nature 352, 589-594.

Barnhart, C.J., Howard, A.D.; Moore, J.M (2009) Long-term precipitation and late-stage valley network formation: Landform simulations of Parana Basin, Mars. J. Geophys. Res. (Planets) 114, E01003.

Beegle, L., et al. (2015), SHERLOC: Scanning habitable environments with Raman & luminescence for organics & chemicals, 2015 IEEE Aerospace Conference, DOI: 10.1109/AERO.2015.7119105.

Bibring, J.-P., et al. (2006), Global mineralogical and aqueous Mars history derived from OMEGA/Mars express data, Science, 312(5772), 400–404.

Bierson, C. J.; Phillips, R. J.; Smith, I. B.; Wood, S. E.; Putzig, N. E.; Nunes, D.; Byrne, S. (2016) Stratigraphy and evolution of the buried $CO_2$ deposit in the Martian south polar cap, Geophys. Res. Lett., 43, 9, 4172-4179

Bishop, J. L.; Fairén, A.G.; Michalski, J.R.; Gago-Duport, L.; Baker, L.L.; Velbel, M. A.; Gross, C.; Rampe, E. B., 2018, Surface clay formation during short-term warmer and wetter conditions on a largely cold ancient Mars, Nat. Astron., 2, 206-213.

Bowen, GJ, BJ Maibauer, MJ Kraus, U Röhl, T Westerhold, A Steimke (2016) Two massive, rapid releases of carbon during the onset of the Palaeocene-Eocene thermal maximum, Nat. Geosci. 8, 44-47.

Brož, P., E. Hauber, I. van de Burgt, V. Špillar, G. Michael (2019), Subsurface Sediment Mobilization in the Southern Chryse Planitia on Mars, https://doi.org/10.1029/2018JE005868.

Carter, J.; Poulet, F.; Bibring, J.-P.; Mangold, N.; Murchie, S. (2013) Hydrous minerals on Mars as seen by the CRISM and OMEGA imaging spectrometers: Updated global view, J. Geophys. Res.: Planets, 118, 831-858.

Carter, J., D. Loizeau, N. Mangold, F. Poulet, J.-P. Bibring (2015) Widespread surface weathering on early Mars: A case for a warmer and wetter climate, Icarus, 248, 373–382.

Cassata, W.S.; Shuster, D.L.; Renne, P.R.; Weiss, B.P., 2012, Trapped Ar isotopes in meteorite ALH 84001 indicate Mars did not have a thick ancient atmosphere, Icarus, 221, 461-465.

Catling, D.C. & Kasting, J.F. (2017) Atmospheric evolution on inhabited and lifeless worlds, Cambridge University Press.

Chambers, J.E. (1999), A hybrid symplectic integrator that permits close encounters between massive bodies. Monthly Notices Royal Astron. Soc. 304, 793-799.

Chassefière, E., & Leblanc, F. (2011) Methane release and the carbon cycle on Mars, Planetary & Space Sci. 59, 207-217

Chassefière, E., Lasue, J., Langlais, B., Quesnel, Y. (2016) Early Mars Serpentinization Derived $CH_4$ Reservoirs and $H_2$ Induced Warming. Meteoritics & Planet. Sci., 51, 2234–2245.

Chastain, B.K.; Chevrier, V. (2007) Methane clathrate hydrates as a potential source for martian atmospheric methane, Planetary and Space Science 55, 1246-1256.

Chevrier, V., Poulet, F., and Bibring, J.-P. (2007) Early geochemical environment of Mars as determined from thermodynamics of phyllosilicates, Nature, 448, 7149, 60-63.

Claire, M.W, et al. (2012) The Evolution of Solar Flux from 0.1 nm to 160 μm, Astrophys. J. 757, article id. 95, 12 pp.

Dethier, D.P. (2001) Pleistocene incision rates in the western United States calibrated using Lava Creek B tephra, Geology, 29, 783-786.

Doran, P.T., et al. (2002) Valley floor climate observations from the McMurdo dry valleys, Antarctica, 1986-2000, J. Geophys. Res. Atmospheres, 107, D24, ACL 13-1, CiteID 4772, DOI 10.1029/2001JD002045.

Edwards, C.S.; Ehlmann, B.L. (2015) Carbon sequestration on Mars. Geology 43, 863-866

Eigenbrode, J.L., R.E. Summons, A. Steele, C. Freissinet, M. Millan, R. Navarro-González, B. Sutter, A.C. McAdam, H.B. Franz, D.P. Glavin, P.D. Archer Jr., P.R. Mahaffy, P.G. Conrad, J.A. Hurowitz, et al. (2018), Organic matter preserved in 3-billion-year-old mudstones at Gale crater, Mars, Science, 360(6393):1096-1101.

Ehlmann, B. L.; Mustard, J. F.; Fassett, C. I.; Schon, S. C.; Head, J. W., III; Des Marais, D. J.; Grant, J. A.; Murchie, S. L. (2008), Clay minerals in delta deposits and organic preservation potential on Mars, Nature Geoscience, 1, 6, 355-358

Ehlmann, B.L., J. F. Mustard, and S. L. Murchie (2010) Geologic setting of serpentine deposits on Mars. Geophys. Res. Lett. , 370:L06201, 2010. doi: 10.1029/2010GL042596.

Ehlmann, B. L., et al. (2011) Subsurface water and clay mineral formation during the early history of Mars, Nature, 479(7371), 53–60.

Etiope, G., Sherwood Lollar, B. (2013) Abiotic Methane on Earth. Rev. Geophys. 51, 276-299.

Etiope, G., & M. Schoell (2014), Abiotic gas: Atypical, but not rare, Elements 10(4), 291-296.
Etiope, Giuseppe; Oehler, Dorothy Z (2019), Methane spikes, background seasonality and non-detections on Mars: A geological perspective, Planetary and Space Science, 168, 52-61.

Fassett, C. I.; Head, J. W. (2005) Fluvial sedimentary deposits on Mars: Ancient deltas in a crater lake in the Nili Fossae region, Geophys. Res. Lett., 32, 14, L14201.

Fassett, C.I., Head, J.W. (2008a). The timing of martian valley network activity: Constraints from buffered crater counting. Icarus 195, 61-89.

Fassett, C. I. and J. W. Head (2008b) Valley network-fed, open-basin lakes on Mars: Distribution and implications for Noachian surface and subsurface hydrology. Icarus, 198, 37-56.

Fassett, C.I., & J. W. Head (2011) Sequence and timing of conditions on early Mars. Icarus 211, 1204-1214, doi: 10.1016/j.icarus.2010.11.014.

Fastook, J. L., and J. W. Head III (2015) Glaciation in the Late Noachian Icy Highlands: Ice accumulation, distribution, flow rates, basal melting, and top-down melting rates and patterns, Planet. Space Sci., 106, 82-98, doi: 10.1016/j.pss.2014.11.028.

Forget, F., Hourdin, F., Talagrand, O., 1998, $CO_2$ Snowfall on Mars: Simulation with a General Circulation Model, Icarus, 131, 302-316.

Forget, F.; et al. (2013), 3D modelling of the early martian climate under a denser CO2 atmosphere: Temperatures and CO2 ice clouds, Icarus 222(1), p. 81-99.

Gaillard, F.; Michalski, J.; Berger, G.; McLennan, S.M.; Scaillet, B. (2013) Geochemical Reservoirs and Timing of Sulfur Cycling on Mars, Space Sci. Rev., 174, 1-4, 251-300.

Gainey, S. R.; Elwood Madden, M. E. (2012) Kinetics of methane clathrate formation and dissociation under Mars relevant conditions, Icarus 218, 513-524.

Gierasch, P. J.; Toon, O. B. (1973) Atmospheric pressure variation and the climate of Mars. J. Atmos. Sci., 30, 1502-1508.

Goudge T.A.; Fassett, C.I.; Head, J.W.; Mustard, J.F.; Aureli, K.L. (2016) Insights into surface runoff on early Mars from paleolake basin morphology and stratigraphy, Geology 44, 419-422.

Goudge, T.A.; Mohrig, D.; Cardenas, B. T.; Hughes, Cory M.; Fassett, C. I. (2018) Stratigraphy and paleohydrology of delta channel deposits, Jezero crater, Mars, Icarus, 301, 58-75.

Grasby, S. E.; Proemse, B. C.; Beauchamp, B. (2014) Deep groundwater circulation through the High Arctic cryosphere forms Mars-like gullies, Geology, 42, 651-654.

Grotzinger, J., D.Y. Sumner, L.C. Kah, K. Stack, S. Gupta, L. Edgar, D. Rubin, K. Lewis, J. Schieber, N. Mangold, R. Milliken, G. Conrad, D. DesMarais, J. Farmer, K. Siebach, F. Calef, J. Hurowitz, S.M. McLennan et al. (2014), A Habitable Fluvio-Lacustrine environment at Yellowknife bay, Gale Crater, Mars, Science 343(6169), 1242777.

Haberle, R. M.; Catling, D. C.; Carr, M. H.; Zahnle, K. J. (2017), The Early Mars Climate System, in The atmosphere and climate of Mars, Edited by R.M. Haberle et al. ISBN: 9781139060172. Cambridge University Press, 2017, 497-525.

Haberle, R. M.; Zahnle, K.; Barlow, N. G. (2018), Warming Early Mars by Impact Degassing of Reduced Greenhouse Gases, 49th Lunar and Planetary Science Conference 19-23 March, 2018, held at The Woodlands, Texas LPI Contribution No. 2083, id.1682.

Halevy, I.; Fischer, W. W.; Eiler, J. M (2011) Carbonates in the Martian meteorite Allan Hills 84001 formed at 18±4°C in a near-surface aqueous environment, Proc. Natl. Acad. Sci., 108, 16895-16899.

Halevy, I., Head, J.W., III (2014) Episodic warming of early Mars by punctuated volcanism. Nat. Geosci. 7, 865-868.

Halevy, I.; Head, J. W., 2017, Atmospheric Mass and the Geologic Record of Water on Mars, 48th Lunar and Planetary Science Conference, held 20-24 March 2017, at The Woodlands, Texas. LPI Contribution No. 1964, id.2519

Haqq-Misra, J. D.; Domagal-Goldman, S. D.; Kasting, P. J.; Kasting, J. F. (2008), A Revised, Hazy Methane Greenhouse for the Archean Earth, Astrobiology, 8, 1127-1137.

Hassler, D.M., et al. (2014) Mars' surface radiation environment measured with the Mars science laboratory's curiosity rover. Science 343, 1244797 .

Hecht, M.H. (2002) Metastability of Liquid Water on Mars. Icarus, 156, 373-386.

Heng, K., & P. Kopparla (2012) On the stability of Super-Earth atmospheres, Astrophys. J. 754, 1.

Hoke, M.R.T. & Hynek, B.M. (2009) Roaming zones of precipitation on ancient Mars as recorded in valley networks, J. Geophys. Res. 114(E8), CiteID E08002.

Hoke, M.R.T.; Hynek, B.M.; Tucker, G.E. (2011) Formation timescales of large Martian valley networks, Earth Planet Sci. Lett. 312, 1-12.

Holo, S. J.; Kite, E. S.; Robbins, S. J. (2018), Mars obliquity history constrained by elliptic crater orientations, Earth Planet.. Sci. Lett., 496, p. 206-214

Hörst, S. M. (2017), Titan's atmosphere and climate, Journal of Geophysical Research: Planets, Volume 122, Issue 3, pp. 432-482.

Howard, A. D.; Moore, J. M.; and Irwin, R. P. (2005) An intense terminal epoch of widespread fluvial activity on early Mars: 1. Valley network incision and associated deposits. J. Geophys. Res. (Planets), 110:E12S14, 2005. doi: 10.1029/2005JE002459.

Hynek, B.M.; Beach, M.; Hoke, M.R. T. (2010) Updated global map of Martian valley networks and implications for climate and hydrologic processes, J. Geophys. Res., 115(E9), CiteID E09008.

Hynek, B. (2016), The great climate paradox of ancient Mars, Geology, 44, 879-880.

Irwin, R. P., A. D. Howard, R. A. Craddock, and J. M. Moore (2005). An intense terminal epoch of widespread fluvial activity on early Mars: 2. Increased runoff and paleolake development. J. Geophys. Res. (Planets) 110, E12S15.

Irwin, R. P.; Tanaka, K. L.; and Robbins, S. J. (2013). Distribution of Early, Middle, and Late Noachian cratered surfaces in the Martian highlands: Implications for resurfacing events and processes. J. Geophys. Res. (Planets) 118, 278-291.

Irwin, R. P., III; Howard, A. D.; and Craddock, R. A. (2008), Fluvial valley networks on Mars, in River

Confluences, Tributaries, and the Fluvial Network, edited by S. Rice, A. Roy, and B. Rhoads, 409–430, John Wiley.

Irwin, R.P., K.W. Lewis, A.D. Howard, J.A. Grant (2015), Paleohydrology of Eberswalde crater, Mars. Geomorphology, https://doi.org/10.1016/j.geomorph.2014.10.012.

Isherwood, R.J.; Jozwiak, L.M.; Jansen, J.C.; Andrews-Hanna, J.C. (2013), The volcanic history of Olympus Mons from paleo-topography and flexural modeling, Earth Planet Sci. Lett., 363, 88-96.

Ivanov, M. A.; Hiesinger, H.; Erkeling, G.; Reiss, D., 2017, Mud volcanism and morphology of impact craters in Utopia Planitia on Mars: Evidence for the ancient ocean, Icarus, 228, 121-140.

Jakosky, B.M.; Edwards, C.S. (2018), Inventory of CO2 available for terraforming Mars, Nature Astronomy, 2, 634-639.

Jakosky, B.M. et al., 2018, Loss of the Martian atmosphere to space: Present-day loss rates determined from MAVEN observations and integrated loss through time, Icarus, 315, 146-157.

Jakosky, B.M., 2019, The CO2 inventory on Mars, Planetary and Space Science, Volume 175, p. 52-59.

Kahre, M.A., Vines, S.K., Haberle, R.M., Hollingsworth, J.L. (2013), The early Martian atmosphere: Investigating the role of the dust cycle in the possible maintenance of two stable climate states, J. Geophys. Res. – Planets, DOI: 10.1002/jgre.20099

Karman, Tijs; Gordon, Iouli E.; van der Avoird, Ad; Baranov, Yury I.; Boulet, Christian; Drouin, Brian J.; Groenenboom, Gerrit C.; Gustafsson, Magnus; Hartmann, Jean-Michel; Kurucz, Robert L.; Rothman, Laurence S.; Sun, Kang; Sung, Keeyoon; Thalman, Ryan; Tran, Ha; Wishnow, Edward H.; Wordsworth, Robin; Vigasin, Andrey A.; Volkamer, Rainer; van der Zande, Wim J., 2019, Update of the HITRAN collision-induced absorption section, Icarus, Volume 328, p. 160-175.

Kasting, JF, & D Catling (2003) Evolution of a habitable planet, Annual Review of Astronomy and Astrophysics 41 (1), 429-463

Kerber, L., Forget, F., Wordsworth, R (2015). Sulfur in the early martian atmosphere revisited: experiments with a 3-D Global Climate Model. Icarus 261, 133-148.

Kite, Edwin S.; Hindmarsh, Richard C. A. (2007) Did ice streams shape the largest channels on Mars? Geophys. Res. Lett., 34, 19, CiteID L19202.

Kite, E. S.; Hovius, N.; Hillier, J. K.; Besserer, J. (2007), Candidate Mud Volcanoes in the Northern Plains of Mars, American Geophysical Union, Fall Meeting 2007, abstract id.V13B-1346

Kite, E.S., Williams, J.-P., Lucas, A., Aharonson, O. (2014) Low palaeopressure of the martian atmosphere estimated from the size distribution of ancient craters. Nat. Geosci. 7, 335–339.

Kite, E.S., Howard, A., Lucas, A., Armstrong, J.C., Aharonson, O., & Lamb, M.P. (2015) Stratigraphy of Aeolis Dorsa, Mars: stratigraphic context of the great river deposits, Icarus, 253, 223-242.

Kite, E.S., Gao P., Goldblatt, C., Mischna M., Mayer D., and Yung Y.L. (2017a) Methane bursts as a trigger for intermittent lake-forming climates on post-Noachian Mars, Nat. Geosci. 10, 737-740.

Kite, E.S., Sneed, J., Mayer, D.P. & Wilson, S.A. (2017b), Persistent or repeated surface habitability on Mars during the Late Hesperian - Amazonian, Geophys. Res. Lett., doi:10.1002/2017GL072660.

Kite, E.S. (2019), Geologic constraints on Early Mars climate, Space Sci. Rev., 215:10.

Kite, E.S., Mayer, D.P., Wilson, S.A., Davis, J.M., Lucas, A.S., & Stucky de Quay, G. (2019), "Persistence of intense, climate-driven runoff late in Mars history," Science Advances, 5(3), eaav7710, doi:10.1126/sciadv.aav7710.

Kitzmann, D. (2016), Revisiting the Scattering Greenhouse Effect of $CO_2$ Ice Clouds, Astrophys. J. Lett., 817, article id. L18, 5 pp.

Klein, F., Grozeva, N.G., and Seewald, J.S. (2019), Abiotic methane synthesis and serpentinization in olivine-hosted fluid inclusions, Proc. Natl. Acad. Sci. 116, 17666-17672.

Komatsu, G.; Kargel, J. S.; Baker, V. R.; Strom, R. G.; Ori, G. G.; Mosangini, C.; Tanaka, K. L. (2000), A Chaotic Terrain Formation Hypothesis: Explosive Outgas and Outlow by Dissociation of Clathrate on Mars, 31st Annual Lunar and Planetary Science Conference, March 13-17, 2000, Houston, Texas, abstract no. 1434

Komatsu, G.; Okubo, C. H.; Wray, J. J.; Ojha, L.; Cardinale, M.; Murana, A.; Orosei, R.; Chan, M. A.; Ormö, J.; Gallagher, R. (2016), Small edifice features in Chryse Planitia, Mars: Assessment of a mud volcano hypothesis, Icarus, 268, 56-75.

Korablev, Oleg; Vandaele, Ann Carine; Montmessin, Franck; Fedorova, Anna A.; Trokhimovskiy, Alexander; Forget, François; Lefèvre, Franck; Daerden, Frank; Thomas, Ian R.; Trompet, Loïc; Erwin, Justin T.; Aoki, Shohei; Robert, Séverine; Neary, Lori; Viscardy, Sébastien; Grigoriev, Alexey V.; Ignatiev, Nikolay I.; Shakun, Alexey; Patrakeev, Andrey; Belyaev, Denis A.; Bertaux, Jean-Loup; Olsen, Kevin S.; Baggio, Lucio; Alday, Juan; Ivanov, Yuriy S.; Ristic, Bojan; Mason, Jon; Willame, Yannick; Depiesse, Cédric; Hetey, Laszlo; Berkenbosch, Sophie; Clairquin, Roland; Queirolo, Claudio; Beeckman, Bram; Neefs, Eddy; Patel, Manish R.; Bellucci, Giancarlo; López-Moreno, Jose-Juan; Wilson, Colin F.; Etiope, Giuseppe; Zelenyi, Lev; Svedhem, Hâkan; Vago, Jorge L.; Acs; NOMAD Science Teams, 2019, No detection of methane on Mars from early ExoMars Trace Gas Orbiter observations, Nature, Volume 568, Issue 7753, p.517-520.

Krasnopolsky, V.A., Maillard, J.P., Owen, T.C. (2004) Detection of methane in the martian atmosphere: evidence for life? Icarus 172, 537-547.

Kreslavsky, M. A.; Head, J. W. (2005) Mars at very low obliquity: Atmospheric collapse and the fate of volatiles, Geophys. Res. Lett., 32, 12, CiteID L12202

Kurahashi-Nakamura, T. and Tajika, E. (2006), Atmospheric collapse and transport of carbon dioxide into the subsurface on early Mars, Geophys. Res. Lett. 33, 18, https://doi.org/10.1029/2006GL027170.

Kurokawa, H., Kurosawa, K., and Usui, T. (2017) A lower limit of atmospheric pressure on Early Mars inferred from nitrogen and argon isotopic compositions, Icarus, doi:10.106/j.icarus.2017.08.020.

Lammer, H., E. et al. (2013) Outgassing History and Escape of the Martian Atmosphere and Water Inventory, Space Sci. Rev., 174, 113–154.

Lapôtre, M. G. A., et al. (2016), Large wind ripples on Mars: A record of atmospheric evolution, Science, 353, 55–58.

Laskar, J., et al. (2004), Long term evolution and chaotic diffusion of the insolation quantities of Mars, Icarus, 170, 343–364.

Lasue, J., Quesnel, Y., Langlais, B., Chassefière, E. (2015) Methane storage capacity of the early martian cryosphere, Icarus 260, 205-214.

Leask, E.K., B. L. Ehlmann M. M. Dundar S. L. Murchie F. P. Seelos (2018) Challenges in the Search for Perchlorate and Other Hydrated Minerals With 2.1-μm Absorptions on Mars, Geophys. Res. Lett., 09 November 2018, https://doi.org/10.1029/2018GL080077

Lewis, Kevin W.; Peters, Stephen; Gonter, Kurt; Morrison, Shaunna; Schmerr, Nicholas; Vasavada, Ashwin R.; Gabriel, Travis, 2019, A surface gravity traverse on Mars indicates low bedrock density at Gale crater, Science, Volume 363, Issue 6426, pp. 535-537.

Lin, Li-Hung; Hall, James; Lippmann-Pipke, Johanna; Ward, Julie A.; Sherwood Lollar, Barbara; Deflaun, Mary; Rothmel, Randi; Moser, Duane; Gihring, Thomas M.; Mislowack, Bianca; Onstott, T. C., 2005, Radiolytic H2 in continental crust: Nuclear power for deep subsurface microbial communities, Geochemistry Geophysics Geosystems, 6(7), Q07003.

Luo, W., Cang, X. & A.D. Howard, (2017) New Martian valley network volume estimate consistent with ancient ocean and warm and wet climate, Nat. Communications 8, 15766

Lunine, Jonathan I.; Atreya, Sushil K. (2008), The methane cycle on Titan, Nature Geoscience 1, 159-164.

Lyons, J.R.; Manning, C.; Nimmo, F. (2005) Formation of methane on Mars by fluid-rock interaction in the crust, Geophys. Res. Lett. 32, CiteID L13201.

Mahaffy, P.R., et al. (2015) The imprint of atmospheric evolution in the D/H of Hesperian clay minerals on Mars. Science 347, 412-414.

Manning, C.V., McKay, C.P., Zahnle, K.J. (2006) Thick and thin models of the evolution of carbon dioxide on Mars. Icarus 180, 38–59.

Manning, C.V., C. Bierson, N.E. Putzig, C.P. McKay (2019) The formation and stability of buried polar $CO_2$ deposits on Mars, Icarus, 317, 509-517.

Matsubara, Y.; Howard, A.D.; Drummond, S.A. (2011) Hydrology of early Mars: Lake basins, J. Geophys. Res., 116, E4, CiteID E04001.

McKay, C. P.; Wharton, R. A., Jr.; Squyres, S. W.; Clow, G. D. (1985) Thickness of ice on perennially frozen lakes, Nature 313, 561-.

McKay, C.P.; Pollack, J.B.; Courtin, R. (1991) The greenhouse and antigreenhouse effects on Titan, Science 253, 1118-1121.

McKay CP, Andersen DT, Pollard WH, Heldmann JL, Doran PT, Fritsen CH, Priscu JC (2005), Polar Lakes, Streams, and Springs as Analogs for the Hydrological Cycle on Mars. pp. 219-233 in Water on Mars and Life, Springer, Berlin, Heidelberg.

Mellon, M.T. (1996) Limits on the $CO_2$ Content of the Martian Polar Deposits, Icarus 124, 268-279.

Milton, D.J. (1974) Carbon Dioxide Hydrate and Floods on Mars, Science, 183, 654-656.

Mischna, Michael A.; Allen, Mark; Richardson, Mark I.; Newman, Claire E.; Toigo, Anthony D. (2011), Atmospheric modeling of Mars methane surface releases, Planetary and Space Science, 59, 227-237.

Mischna, M.A., Baker, V., Milliken, R., Richardson, M., Lee, C. (2013) Effects of obliquity and water vapor/trace gas greenhouses in the early martian climate. J. Geophys. Res. (Planets) 118, 560-576.

Moores, J.E., et al. (2019), The methane diurnal variation and micro‐seepage flux at Gale crater, Mars as constrained by the ExoMars Trace Gas Orbiter and Curiosity observations, Geophysical Research Letters, DOI: 10.1029/2019GL083800

Mousis, O., et al. (2013) Volatile Trapping in Martian Clathrates. Space Sci. Rev. 174, 213-250.

Nair, H.; Allen, M.; Anbar, A.D.; Yung, Y.L.; Clancy, R.T. (1994) A photochemical model of the martian atmosphere, Icarus 111, 124-150.

Nair, H.; Summers, M.E.; Miller, C.E.; Yung, Y.L. (2005) Isotopic fractionation of methane in the martian atmosphere, Icarus 175, 32-35.

Nakamura, T.; Tajika, E. (2001) Stability and evolution of the climate system of Mars, Earth, Planets and Space, 53, 851-859.

Nakamura, T.; Tajika, E., (2002) Stability of the Martian climate system under the seasonal change condition of solar radiation, J. Geophys. Res. (Planets), 107(E11), 4-1, CiteID 5094.

Nakamura, T.; Tajika, E. (2003) Climate change of Mars-like planets due to obliquity variations: implications for Mars, Geophys. Res. Lett., 30, 18-1, CiteID 1685.

Neveu, Marc; Hays, Lindsay E.; Voytek, Mary A.; New, Michael H.; Schulte, Mitchell D. (2018), The Ladder of Life Detection, Astrobiology, 18, 1375-1402.

Oehler, Dorothy Z.; Allen, Carlton C. (2010), Evidence for pervasive mud volcanism in Acidalia Planitia, Mars, Icarus, Volume 208, Issue 2, p. 636-657.

Oehler, Dorothy Z.; Etiope, Giuseppe, 2017, Methane Seepage on Mars: Where to Look and Why, Astrobiology, 17, 12, 2017,1233-1264

Oze, C.; Sharma, M. (2005) Have olivine, will gas: Serpentinization and the abiogenic production of methane on Mars, Geophys. Res. Lett., 32, 10, CiteID L10203

Oze, C.; Jones, L. C.; Goldsmith, J. I.; Rosenbauer, R. J. (2012), Differentiating biotic from abiotic methane genesis in hydrothermally active planetary surfaces, Proceedings of the National Academy of Sciences, 109, 9750-9754.

Pan, L.; Ehlmann, B. L. (2014), Phyllosilicate and hydrated silica detections in the knobby terrains of Acidalia Planitia, northern plains, Mars, Geophys. Res. Lett., 41, 1890-1898.

Parmentier, E.M., Zuber, M.T. (2007) Early evolution of Mars with mantle compositional stratification or hydrothermal crustal cooling. JGR (Planets) 112, E02007.

Phillips, R.J., B.J. Davis, K.L. Tanaka et al. (2011) Massive $CO_2$ ice deposits sequestered in the South polar layered deposits of Mars. Science 332, 838–841.

Prieto-Ballesteros, O., et al. (2006) Interglacial clathrate destabilization on Mars: Possible contributing source of its atmospheric methane. Geology 34, 149.

Putzig, Nathaniel E.; Smith, Isaac B.; Perry, Matthew R.; Foss, Frederick J.; Campbell, Bruce A.; Phillips, Roger J.; Seu, Roberto (2018) Three-dimensional radar imaging of structures and craters in the Martian polar caps, Icarus, 308, 138-147.

Ramirez, R. M.; Kopparapu, R.; Zugger, M. E.; Robinson, T.D.; Freedman, R.; Kasting, J. F. (2014) Warming early Mars with $CO_2$ and $H_2$, Nature Geoscience, 7, 59-63.

Ramirez, R. M. (2017) A warmer and wetter solution for early Mars and the challenges with transient warming, Icarus, 297, 71-82.

Richardson, M.I., Toigo, A.D., Newman, C.E. (2007) PlanetWRF: a general purpose, local to global numerical model for planetary atmospheric and climate dynamics. J. Geophys. Res. 112, E09001.

Robbins, S. J., B. M. Hynek, R. J. Lillis, and W. F. Bottke (2013) Large impact crater histories of Mars: The effect of different model crater age techniques. Icarus 225, 173-184.

Roe, G. (2009) Feedbacks, timescales, and seeing red, Ann. Rev. Earth Planet. Sci. 37, 93-115

Rosenberg, E.N.; Palumbo, A.M.; Cassanelli, J.P.; Head, J.W.; Weiss, D.K. (2019), The volume of water required to carve the martian valley networks: Improved constraints using updated methods, Icarus, 317, 379-387.

Sagan, C.; Toon, O. B., Gierasch, P. (1973) Climatic Change on Mars, Science, 181, 1045-1049.

Sagan, C. (1977), Reducing greenhouses and the temperature history of Earth and Mars, Nature 269, 224 - 226; doi:10.1038/269224a0

Schorghofer, N. (2008), Temperature response of Mars to Milankovitch cycles, Geophys. Res. Lett., 35, L18201

Shaw, J., Healy, T. (1980) Morphology of the Onyx River system, McMurdo sound region, Antarctica. N. Z. J. Geol. Geophys. 23 (2), 223–238.

Skinner, J.A., & Tanaka, K.L. (2007) Evidence for and implications of sedimentary diapirism and mud volcanism in the southern Utopia highland–lowland boundary plain, Mars, Icarus, 186, 41-59.

Sloan, E.D., & Koh, C.A. (2008) Clathrate Hydrates of Natural Gases (3rd Edition), CRC Press.

Som, S. M.; Montgomery, D. R.; Greenberg, H. M. (2009) Scaling relations for large Martian valleys, J. Geophys. Res. - Planets 114(E2), CiteID E02005.

Soto, A.; Mischna, M.; Schneider, T.; Lee, C.; Richardson, M.I. (2015), Martian atmospheric collapse: Idealized GCM studies, Icarus, 250, 553-569.

Stern, L.A., Circone, S., Kirby, S.H., Durham, W.B. (2003) Temperature, pressure, and compositional effects on anomalous or "self" preservation of gas hydrates. Can. J. Phys. 81, 271-283.

Steakley, K., Kahre, M., Haberle, R., & Zahnle, K. (2019), EPSC/DPS Joint Meeting, abstract number EPSC-DPS2019-509-1.

Summers, M.E.; Lieb, B. J.; Chapman, E.; Yung, Y.L. (2002) Atmospheric biomarkers of subsurface life on Mars, Geophys. Res. Lett. 29, 24-1, CiteID 2171.

Summons, R.E., et al. (2011) Preservation of martian organic and environmental records: Final report of the Mars biosignature working group, Astrobiology, 11(2): doi:10.1089/ast.2010.0506.

Sun, V.Z.; Milliken, R.E. (2015) Ancient and recent clay formation on Mars as revealed from a global survey of hydrous minerals in crater central peaks, J. Geophys. Res. Planets, 120, 2293-2332.

Sutter, B.; McAdam, A. C.; Mahaffy, P. R.; Ming, D. W.; Edgett, K. S.; Rampe, E. B.; Eigenbrode, J. L.; Franz, H. B.; Freissinet, C.; Grotzinger, J. P.; et al. (2017), Evolved gas analyses of sedimentary rocks and eolian sediment in Gale Crater, Mars: Results of the Curiosity rover's sample analysis at Mars instrument from Yellowknife Bay to the Namib Dune, J. Geophys. Res.: Planets, 122, 2574-2609

Tarnas, J. D.; Mustard, J. F.; Sherwood Lollar, B.; Bramble, M. S.; Cannon, K. M.; Palumbo, A. M.; Plesa, A.-C., 2018, Radiolytic H2 production on Noachian Mars: Implications for habitability and atmospheric warming, Earth & Planetary Science Letters, 502, 133-145.

Tarnas, J. D.; Mustard, J. F.; Sherwood Lollar, B.; Cannon, K. M.; Palumbo, A. M.; Plesa, A.-C.; Bramble, M. S., 2019, 50th Lunar and Planetary Science Conference, held 18-22 March, 2019 at The Woodlands, Texas. LPI Contribution No. 2132, id.2029.

Tian, F., Kasting, J.F., & Solomon, S.C. (2009) Thermal escape of carbon from the early Martian atmosphere, Geophys. Res. Lett., DOI: 10.1029/2008GL036513

Tobie, G., Lunine, J.I. and Sotin, C. (2006), Episodic outgassing as the origin of atmospheric methane on Titan, Nature 440, 61-64.

Toigo, A.D., Lee, C., Newman, C.E., Richardson, M.I. (2012) The impact of resolution on the dynamics of the martian global atmosphere: varying resolution studies with the MarsWRF GCM. Icarus 221, 276–288.

Tosca, N.J., Imad A. M. Ahmed, B. M. Tutolo, A. Ashpitel & J. A. Hurowitz (2018), Magnetite authigenesis and the warming of early Mars, Nat. Geosci., 11, 635–639.

Touma, J.; Wisdom, J. (1993), The chaotic obliquity of Mars, Science (ISSN 0036-8075), 259, 1294-1297.

Trainer, M. G.; Pavlov, A. A.; Dewitt, H. L.; Jimenez, J. L.; McKay, C. P.; Toon, O. B.; Tolbert, M. A. (2006), Inaugural Article: Organic haze on Titan and the early Earth, Proceedings of the National Academy of Sciences, 103, 18035-18042.

Turbet, M., F. Forget, J. Leconte, B. Charnay & G. Tobie (2017) $CO_2$ condensation is a serious limit to the deglaciation of Earth-like planets, Earth Planet.. Sci. Lett., 476, 11-21.

Turbet, M.; Tran, H.; Pirali, O.; Forget, F.; Boulet, C.; Hartmann, J.-M. (2019) Far infrared measurements of absorptions by $CH_4$+$CO_2$ and $H_2$+$CO_2$ mixtures and implications for greenhouse warming on early Mars, Icarus, 321, 189-199.

Vasavada, A. R.; Milavec, T. J.; Paige, D. A. (1993), Microcraters on Mars - Evidence for past climate variations, J. Geophys. Res. (ISSN 0148-0227), vol. 98, no. E2, 3469-3476.

Vasavada, A. R. (2017), Our changing view of Mars, Physics Today, 70, 3, 34-41.

Vandaele, A. C.; Neefs, E.; Drummond, R.; Thomas, I. R.; Daerden, F.; Lopez-Moreno, J.-J.; Rodriguez, J.; Patel, M. R.; Bellucci, G.; Allen, M.; et al., (2015), Science objectives and performances of NOMAD, a spectrometer suite for the ExoMars TGO mission, Planet. Space Sci., 119, 233-249.

Ward, M. K.; Pollard, W. H. (2018), A hydrohalite spring deposit in the Canadian high Arctic: A potential Mars analogue, Earth Planet.. Sci. Lett., 504, 126-138.

Warren, A.O., Kite E.S., Williams, J.-P., & Horgan, B. (2019), Through the thick and thin: New constraints on Martian paleopressure history 3.8-4 Ga from small exhumed craters," Journal of Geophysical Research – Planets, doi.org/10.1029/2019JE006178.

Webster, C. R.; Mahaffy, P. R.; Atreya, S. K.; Moores, J. E.; Flesch, G. J.; Malespin, C.; McKay, C. P.; Martinez, G.; Smith, C. L.; Martin-Torres, J.; et al. (2018) Background levels of methane in Mars' atmosphere show strong seasonal variations, Science, 360, 1093-1096.

Wordsworth, R.; Forget, F.; Millour, E.; Head, J. W.; Madeleine, J.-B.; Charnay, B. (2013), Global modelling of the early martian climate under a denser CO2 atmosphere: Water cycle and ice evolution, Icarus, 222, 1-19.

Wordsworth, R., 2015, Atmospheric Heat Redistribution and Collapse on Tidally Locked Rocky Planets, The Astrophysical Journal, Volume 806, Issue 2, article id. 180, 15 pp.

Wordsworth, R.; Kerber, L.; Pierrehumbert, R.; Forget, F.; Head, J.W. (2015), Comparison of warm and wet and cold and icy scenarios for early Mars in a 3-D climate model, JGR: Planets, 120, 1201-1219.

Wordsworth, R. (2016), The Climate of Early Mars, Annual Reviews of Earth and Planetary Sciences, 44, 381-408, https://doi.org/10.1146/annurev-earth-060115-012355.

Wordsworth, R., et al (2017) Transient reducing greenhouse warming on early Mars, Geophys. Res. Lett. 44, 665-671.

Wray, James J.; Ehlmann, Bethany L. (2011), Geology of possible Martian methane source regions, Planetary and Space Science, Volume 59, Issue 2-3, p. 196-202.

Yen, A. S.; Ming, D. W.; Vaniman, D. T.; Gellert, R.; Blake, D. F.; Morris, R. V.; Morrison, S. M.; Bristow, T. F.; Chipera, S. J.; Edgett, K. S.; Treiman, A. H.; Clark, B. C.; Downs, R. T.; Farmer, J. D.; et al. (2017), Multiple stages of aqueous alteration along fractures in mudstone and sandstone strata in Gale Crater, Mars, Earth Planet. Sci. Lett., 471, 186-198.

Zabrusky, K.; Andrews-Hanna, J.C.; Wiseman, S. (2012), Reconstructing the distribution and depositional history of the sedimentary deposits of Arabia Terra, Mars, Icarus 220, 311-330.

Zahnle, K., Freedman, R.S., and Catling, D.C. (2011) Is there methane on Mars? Icarus 212, 493-503.

Zahnle, K. & Catling, D. (2017) The Cosmic Shoreline: The Evidence that Escape Determines which Planets Have Atmospheres, and what this May Mean for Proxima Centauri B, Astrophys. J., 843:122 .

Zahnle, K., & Catling, D. (2019) The paradox of Mars methane, Ninth International Conference on Mars, abstract #6132.